\documentstyle[emulateapj,psfig]{article}

\makeatletter

\newenvironment{inlinefigure}{%
\def\@captype{figure}%
\noindent\begin{minipage}{0.999\linewidth}\begin{center}}
{\end{center}\end{minipage}\smallskip}
\makeatother

\lefthead{Barger et al.}

\begin{document}
\title{Optical and Infrared Properties of 
the 2~Ms {\it Chandra} Deep Field-North X-ray Sources
\altaffilmark{1,2}}
\author{
A.~J.~Barger,$\!$\altaffilmark{3,4,5}
L.~L.~Cowie,$\!$\altaffilmark{5}
P.~Capak,$\!$\altaffilmark{5}
D.~M.~Alexander,$\!$\altaffilmark{6} 
F.~E.~Bauer,$\!$\altaffilmark{6}
E.~Fernandez,$\!$\altaffilmark{5,7}
W.~N.~Brandt,$\!$\altaffilmark{6}
G.~P.~Garmire,$\!$\altaffilmark{6}
A.~E.~Hornschemeier$\!$\altaffilmark{8,9}
}

\altaffiltext{1}{Based in part on data obtained at the Subaru Telescope,
which is operated by the National Astronomical Observatory of Japan.}
\altaffiltext{2}{Based in part on data obtained at the W. M. Keck
Observatory, which is operated as a scientific partnership among the
the California Institute of Technology, the University of
California, and NASA and was made possible by the generous financial
support of the W. M. Keck Foundation.}
\altaffiltext{3}{Department of Astronomy, University of 
Wisconsin-Madison, 475 North Charter Street, Madison, WI 53706.}
\altaffiltext{4}{Department of Physics and Astronomy,
University of Hawaii, 2505 Correa Road, Honolulu, HI 96822.}
\altaffiltext{5}{Institute for Astronomy, University of Hawaii,
2680 Woodlawn Drive, Honolulu, HI 96822.}
\altaffiltext{6}{Department of Astronomy \& Astrophysics,
525 Davey Laboratory, The Pennsylvania State University,
University Park, PA 16802.}
\altaffiltext{7}{Department of Physics, 
New Mexico Institute of Mining and Technology,
801 Leroy Place, Socorro, NM, 87801.}
\altaffiltext{8}{Chandra Fellow.}
\altaffiltext{9}{Department of Physics and Astronomy, Johns
Hopkins University, 3400 North Charles Street, Baltimore, MD
21218.}

\slugcomment{The Astronomical Journal, in press (August 2003)}

\begin{abstract}
We present an optical and near-infrared catalog for
the X-ray sources in the $\approx2$~Ms {\it Chandra} observation 
of the Hubble Deep Field-North region. We have high-quality
multicolor imaging data for all 503 X-ray point sources in the 
X-ray-selected catalog and reliable spectroscopic redshifts for 
284. We spectroscopically identify six high-redshift ($z>1$) 
type~II quasars ($L_{2-8~{\rm keV}}>10^{44}$~ergs~s$^{-1}$) 
in our sample. Our spectroscopic completeness for the $R\le 24$ 
sources is 87\%. The spectroscopic redshift distribution shows two 
broad redshift spikes that have clearly grown over those originally 
seen in the $\approx1$~Ms exposure. The spectroscopically identified 
extragalactic sources already comprise 75\% of the measured $2-8$~keV 
light. Redshift slices versus $2-8$~keV flux show that
an impressive 54\% of the measured $2-8$~keV light arises from 
sources at $z<1$ and 68\% from sources at $z<2$. Thus, major 
accretion onto supermassive black holes has occurred since the 
universe was half its present age. 

We use seven broadband colors and a Bayesian photometric 
redshift estimation code to obtain photometric redshifts for
the X-ray sources. We find that the photometric redshifts 
are within 25\% of the spectroscopic redshifts for
94\% of the non-broad-line sources with both
photometric and spectroscopic measurements.
The photometrically identified sources 
show a smooth continuation of the spectroscopically identified 
sources to redder $R-HK'$ color with increasing redshift, 
consistent with the galaxy tracks of evolved bulge-dominated
galaxies. Fourteen have colors $R-HK'>5.7$ that 
would classify them as Extremely Red Objects (EROs). The
photometric redshifts of these EROs are all between 
$z\sim 1.5$ and $z\sim 2.5$. 

We use our wide wavelength coverage to determine rest-frame 
colors for the X-ray sources with spectroscopic or photometric 
redshifts. We find that many of the X-ray sources
have the rest-frame colors of evolved red galaxies and that 
there is very little evolution in these colors with redshift.
We also determine absolute magnitudes and find that many of 
the non-broad-line sources are more luminous than $M_I^\ast$,
even at high redshifts. 
We therefore infer that deep X-ray observations 
may provide an effective way of locating $M^\ast$ galaxies with 
colors similar to present-day early-type galaxies to high 
redshifts.
\end{abstract}


\section{Introduction}
\label{secintro}

High-energy X-rays can penetrate extremely large column
densities of gas and dust; hence, hard X-ray surveys are our 
best current method for obtaining the least biased samples
of active galactic nuclei (AGNs). 
Ultradeep surveys with the {\it Chandra X-ray Observatory}
detect the faintest X-ray sources in both the
soft ($0.5-2$~keV) and hard ($2-8$~keV) X-ray energy bands
and have revealed large numbers of AGNs that were missed in 
optical surveys. In the present paper we describe the optical
and near-infrared properties of the 503 X-ray point-sources 
detected in the $\approx2$~Ms {\it Chandra} exposure of the region 
around the Hubble Deep Field-North called the {\it Chandra} 
Deep Field-North (CDF-N). This exposure is the deepest X-ray 
image to date.

\markcite{barger02}Barger et al.\ (2002) presented
a similar catalog of the 370 X-ray point-sources detected
in the $\approx1$~Ms CDF-N exposure 
(\markcite{brandt01}Brandt et al.\ 2001).
They had very complete redshift identifications 
(78\%) for the $R\le 24$ galaxy sources in a $10'$ radius 
around the approximate X-ray image center and found 
spectroscopic evidence for large-scale structure in the 
field, which could account for a part of the field-to-field 
variation seen in the X-ray number counts 
(e.g., \markcite{cowie02}Cowie et al.\ 2002;
\markcite{yang03}Yang et al.\ 2003). Such
large-scale structure is also seen in the spectroscopic
redshift distribution (\markcite{hasinger02}Hasinger 2002;
\markcite{gilli03}Gilli et al.\ 2003)
of the X-ray sources in the $\approx 1$~Ms exposure of the 
{\it Chandra} Deep Field-South 
(\markcite{giacconi02}Giacconi et al.\ 2002).

\markcite{cowie03}Cowie et al.\ (2003) constructed rest-frame
$2-8$~keV luminosity functions versus redshift for the 
AGNs in a number of {\it Chandra} (including the 1~Ms CDF-N), 
{\it ROSAT}, and {\it ASCA} deep fields. 
These authors found that at $z=0.1-1$, 
most of the $2-8$~keV energy density arises in sources with
luminosities in the $10^{42}-10^{44}$~ergs~s$^{-1}$ range.
They also showed that the number density of sources
in this luminosity range rises with decreasing redshift,
while the number density of higher luminosity sources exhibit 
the well-known peak at $z=1.5-3$.
\markcite{cowie03}Cowie et al.\ (2003) argued that
the dominant supermassive black hole formation occurred at 
recent times in sources with low accretion, 
rather than at earlier times in more X-ray luminous sources 
with high accretion.

The 1~Ms {\it Chandra} surveys have also detected apparently 
normal galaxies with X-ray-to-optical flux ratios lower than 
those of AGNs, e.g., $\log(f_{0.5-2~{\rm keV}}/f_R)\lesssim-1$
(\markcite{barger02}Barger et al.\ 2002;
\markcite{rosati02}Rosati et al.\ 2002). 
The majority of X-ray sources with 
$-1\lesssim\log(f_{0.5-2~\rm{keV}}/f_R)\lesssim-2$ have been shown
to be consistent with infrared and radio-emitting starburst
galaxies (\markcite{alex02a}Alexander et al.\ 2002a;
\markcite{bauer02}Bauer et al.\ 2002), while those with even 
lower X-ray-to-optical flux ratios are generally consistent 
with quiescent galaxies with low X-ray luminosities 
($L_{0.5-2~{\rm keV}}\lesssim 10^{41}$~ergs~s$^{-1}$;
\markcite{horn03}Hornschemeier et al.\ 2003).
These apparently normal galaxies are distinct from the X-ray
luminous, ``optically normal'' galaxies that have been
discovered in {\it Chandra} and {\it XMM-Newton} surveys
(e.g., \markcite{barger01b}Barger et al.\ 2001b;
\markcite{comastri02}Comastri et al.\ 2002).
The latter sources show no high ionization signatures of 
AGN activity in their optical spectra, and their hard to soft 
X-ray flux ratios suggest they are highly absorbed systems
whose column densities could effectively extinguish the
ultraviolet, optical, and near-infrared continua from the 
AGNs. Alternatively, \markcite{moran02}Moran, Filippenko, \&
Chornock (2002) propose that, since the entire
galaxy often falls within the spectrograph slit for these
distant systems, the host galaxy light from stars and HII 
regions could be overwhelming the emission-line signatures 
of the AGN activity.

The 1~Ms exposure of the CDF-N has since been extended to
a second megasecond (\markcite{alex03}Alexander et al.\ 2003),
enabling the detection of increasing numbers of
faint AGNs and apparently normal galaxies.
In this paper we present multicolor imaging and 
optical spectroscopy of the X-ray sources 
detected in the $\approx 2$~Ms CDF-N exposure. We
use these data to characterize the redshift distribution 
and properties of the faint X-ray sources.

We take $H_0=65\ h_{65}$~km~s$^{-1}$~Mpc$^{-1}$ and use an
$\Omega_{\rm M}=1/3$, $\Omega_\Lambda=2/3$ cosmology.

\section{X-ray Sample}
\label{secsample}

\markcite{alex03}Alexander et al.\ (2003; hereafter, A03) 
presented the 1.945~Ms CDF-N X-ray images, along with details
of the observations, data reduction, and technical analysis.
A03 merged their point-source lists in seven X-ray 
bands, $0.5-8$~keV (full band), $0.5-2.0$~keV (soft band; SB), 
$2-8$~keV (hard band; HB), $0.5-1$~keV (SB1), $1-2$~keV (SB2),
$2-4$~keV (HB1), and $4-8$~keV (HB2), into a catalog (Tables~3A 
and 3B in A03) of 503 significantly detected point sources over 
an area of about 460~arcmin$^2$. Near the aim point, the data 
reach limiting fluxes of $\approx 2.5\times 10^{-17}$ ($0.5-2$~keV) 
and $\approx 1.4\times 10^{-16}$~ergs~cm$^{-2}$~s$^{-1}$ ($2-8$~keV).
In Table~A1 of the Appendix we present optical magnitudes and
spectroscopic measurements, where available, for the full X-ray
point source catalog of A03. 

\section{Optical/Near-infrared Imaging Data}
\label{secimaging}

The optical imaging data consist of Johnson $B$, Johnson $V$,
Cousins $R$, Cousins $I$, and Sloan $z'$ observations obtained
with the Subaru prime-focus camera Suprime-Cam
(\markcite{miyazaki02}Miyazaki et al.\ 2002)
on the Subaru 8.2~m telescope during
February-April of 2001 and 2002. The Suprime-Cam observations 
and reductions are described in \markcite{capak03}Capak et al.\ (2003a), 
where catalogs of the entire sample of galaxies and stars in the 
field can be found. $HK'$ band observations were obtained with 
the Quick Infrared Camera QUIRC
(\markcite{hodapp96}Hodapp et al.\ 1996) on the University of
Hawaii 2.2~m telescope. The notched $HK'$ filter has a central
wavelength of 1.8~$\mu$m and covers the longer wavelength region
of the $H$ band and the shorter wavelength region of the $K$ band
(roughly the $K'$ filter). Because of its broad bandpass and
low sky background, this filter is extremely fast and is roughly
twice as sensitive as the $H$, $K'$, or $K_S$ filters.
\markcite{barger99}Barger et al.\ (1999) found the empirical
relation $HK'-K=0.13+0.05(I-K)$ to convert between the $HK'$
and $K$ bands; for galaxies at most redshifts, $HK'-K\approx 0.3$.
$U$ band observations were
obtained with the MOSAIC prime-focus camera
(\markcite{jacoby98}Jacoby et al.\ 1998;
\markcite{muller98}Muller et al.\ 1998;
\markcite{wolfe98}Wolfe et al.\ 1998)
on the Kitt Peak National Observatory 4~m telescope.
The $HK'$ and $U$ band data are described
in \markcite{capak03}Capak et al.\ (2003a). 

The X-ray catalog of A03 has extremely good positional
accuracy, which simplifies the problem of matching the
X-ray sources to their optical counterparts. For individual 
X-ray sources, the X-ray positional uncertainties are expected 
primarily to reflect the centroiding accuracies and the overall 
distortions in the X-ray data, and are dependent on the off-axis 
angles and the detected counts. For each source, A03 gives an 
$\approx 80$\% confidence X-ray positional uncertainty in 
column~[4] of their Table~3A. 
The largest such positional uncertainty is just under $1.9''$
for a small number of sources at the outside edge of the
{\it Chandra} field, while about 90\% of the sources have
positional uncertainties less than $1.25''$.

The absolute astrometry of both the X-ray catalog of A03 
and the optical catalog of \markcite{capak03a}Capak et al.\ (2003a) 
have been matched to the VLA 20~cm catalog of
\markcite{richards00}Richards (2000).
We tested the accuracy of the individual
optical source positions relative to the VLA 20~cm
positions for the 20~cm sources lying in the CDF-N region
and found a $1\sigma$ deviation of $0.22''$.
Eighty percent of the optical sources with 20~cm counterparts
lie within $0.5''$ of the radio positions. Since the
radio and optical morphologies may differ, these numbers should
represent an upper bound on the accuracy of the optical positions,
which are generally more accurate than the X-ray positions.

Given the X-ray positional uncertainties, we identified
an X-ray source with an optical counterpart if an optical
source brighter than $R=26.4$ ($5\sigma$) or $z'=25.4$ ($5\sigma$) 
is within $2''$ of the X-ray position. If more than one such 
optical counterpart is within the search radius, then we 
identified the X-ray source with the nearest optical neighbor. 
Using this criterion, there are 429 
X-ray sources with optical 
counterparts in the A03 sample, 278 of which have magnitudes
$R\le 24$. From
\markcite{capak03a}Capak et al.\ (2003a), the surface density of
galaxies and stars in the CDF-N region to $R=24$ is approximately
16.6~arcmin$^{-2}$. Thus, with a search radius of $2''$ and a sample
of 503 sources, we may expect a random field contamination of about
30 sources to $R=24$, or roughly 10\%. (A more sophisticated
analysis allowing for clustering would yield a slightly
higher value, e.g., \markcite{mcmahon02}McMahon et al.\ 2002.)
This contamination rises at fainter magnitudes. For $R=24-26$, 
where the field surface density is 50.1~arcmin$^{-2}$, we may 
expect that about 34 of the 140 optical
identifications, or roughly 25\%, are chance projections, based
on the 225
sources that are not identified with $R\le24$ sources.
The high contamination rate at faint optical magnitudes emphasizes
the need for the highest possible spatial accuracy in the X-ray sample.
If we restrict to the 403 sources where the X-ray positional
uncertainties are less than $1''$, and if we use a smaller search
radius of $1''$ to match to the optical sources, then we find that
307 of the X-ray sources have $R<26$ counterparts, and less than
10\% of these will be random field sources.

Seven sources lie within the envelopes of bright galaxies
but are separated by more than $2''$ and less than $5''$
from the nuclear positions. Since these may be sources lying
within the galaxies, we have placed the optical positions of
these off-axis X-ray sources
(sources 121, 197, 270, 316, 404, 410, and 479)
at the bright galaxy nuclei.
However, the number of examples is sufficiently small that
some of these may well just be chance projections of sources lying
behind the bright galaxies.
Four of these seven sources lie within $2.5''$ of the X-ray
positions, the faintest having $R=21.25$, while two lie within
$4''$ ($19<R<20.25$) and one within $5''$ ($R<19$).
With 503 sources in our X-ray sample,
we expect three sources to lie within $4''$ of an $R<20.25$
galaxy and three sources to lie within $2.5''$ of an $R<21.25$
galaxy, based on the surface density of $R$ band selected sources
in the CDF-N region (\markcite{capak03a}Capak et al.\ 2003a).

Since many of the X-ray sources have positions which are much
more accurate than our chosen $2''$ search radius, we can examine 
how accurate the A03 estimates of the X-ray positional
uncertainties are, and how reliable our counterpart identifications
are, by plotting the distribution of $R\le 24$ counterparts versus
radial X-ray-optical separation in arcseconds for three different 
ranges of X-ray positional uncertainty. In Figures~\ref{fig1}a, 1b, 
and 1c we plot the $R\le 24$ sources with X-ray positional 
uncertainties $0''-0.5''$, $0.5''-1''$, and $1''-1.5''$, 
respectively. Dashed lines show how many random $R\le24$ sources 
are expected as a function of radial X-ray-optical separation,
assuming a uniform spatial distribution.
If the A03 X-ray positional uncertainties are accurate, then we
should see tighter distributions when we look at smaller X-ray
positional uncertainty ranges, and this is indeed what we see.
For the two highest accuracy bins,
random field contamination is negligible and
less than 20\% of the counterparts lie beyond the 80~\% X-ray 
positional uncertainty limits in both cases.

%
%
\begin{inlinefigure}
\psfig{figure=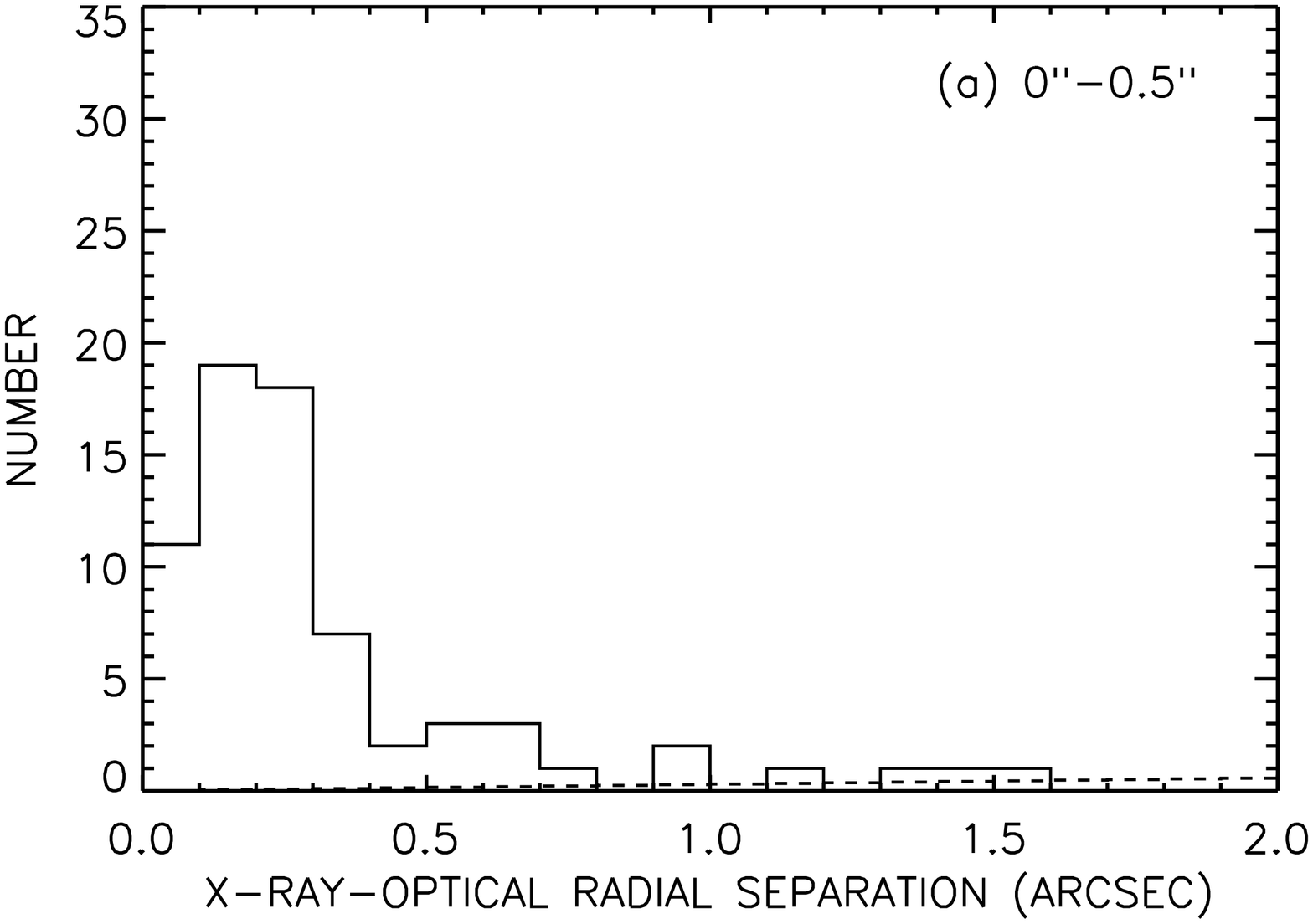,angle=90,width=3.5in}
\psfig{figure=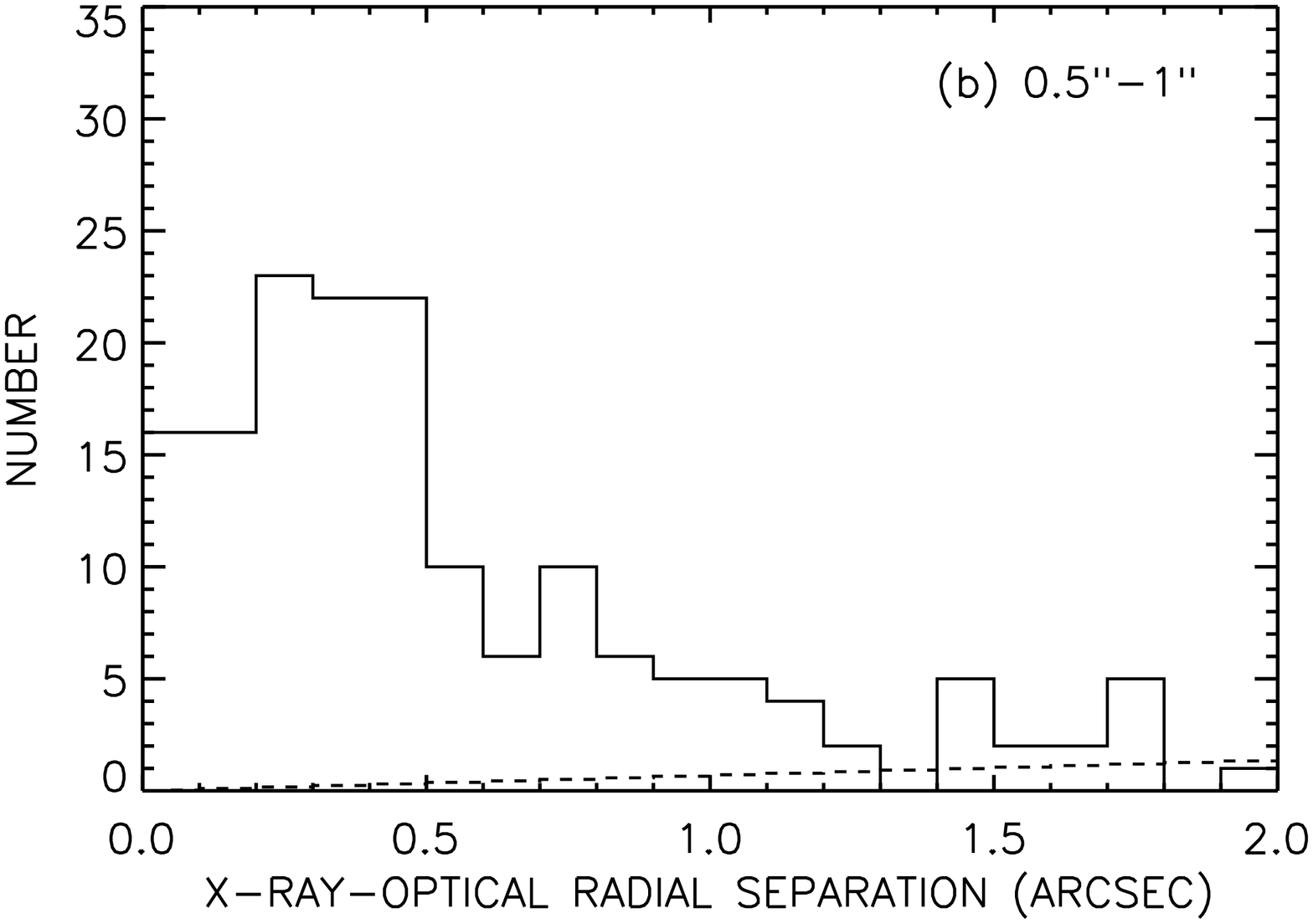,angle=90,width=3.5in}
\psfig{figure=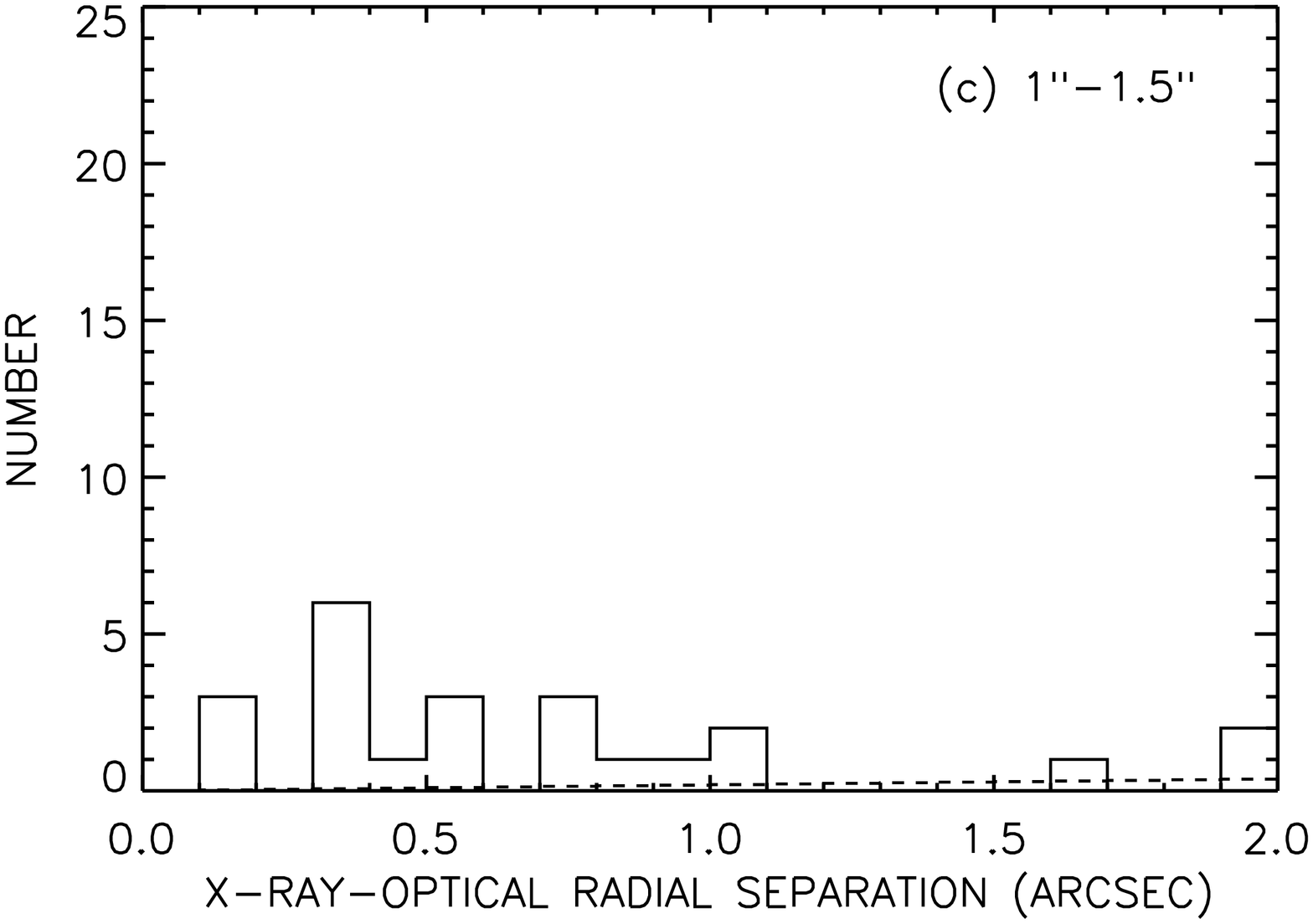,angle=90,width=3.5in}
\vspace{6pt}
\figurenum{1}
\caption{
Optical counterpart distribution vs. radial X-ray-optical
separation in arcseconds for the $R\le 24$ sources
with X-ray positional uncertainties (a) $0''-0.5''$,
(b) $0.5''-1''$, and (c) $1''-1.5''$. Dashed lines show how
many random $R\le 24$ sources are expected as a function of
the radial separation.
\label{fig1}
}
\addtolength{\baselineskip}{10pt}
\end{inlinefigure}

For the sources with optical counterparts, we measured the 
magnitudes centered on the optical positions. (The optical 
positions are set to the peak of the optical emission.)
For the 67 X-ray sources with optically
fainter counterparts, we centered on the X-ray positions.
In all cases we measured the optical magnitudes in $3''$ diameter
apertures and corrected these to total magnitudes using an average
offset in color. For the $R$ band we also give isophotal magnitudes.
Further details may be found in 
\markcite{capak03a}Capak et al.\ (2003a).

Table~A1 of the Appendix gives $B$ and $V$ magnitudes in 
the Johnson system, $R$ and $I$ magnitudes in the Kron-Cousins 
system, $HK'$ magnitudes in the Wainscoat-Cowie system
(\markcite{wains92}Wainscoat \& Cowie 1992), and
$U$ and $z'$ magnitudes in the AB magnitude system.
An AB magnitude is defined by $m_{AB}=-2.5\log f_\nu - 48.60$,
where $f_\nu$ is the flux of the source in units of
ergs~cm$^{-2}$~s$^{-1}$~Hz$^{-1}$.
The $5\sigma$ magnitude limits of the images in the above systems
are 27.1 ($U$), 27.0 ($B$), 26.8 ($V$), 26.4 ($R$), 25.1 ($I$), 
25.4 ($z'$), and 20.5 ($HK'$).

We show $z'$ band thumbnail images of the CDF-N sources in Figure~A1
of the Appendix. In a few cases there is complex structure in
the neighborhood of the X-ray position that complicates the photometry
and counterpart identification. We have noted in Table~A1 the sources 
for which this might be a problem. 

\section{X-ray-to-Optical Flux Ratios}
\label{secopt}

\begin{figure*}[tbh]
\centerline{\psfig{figure=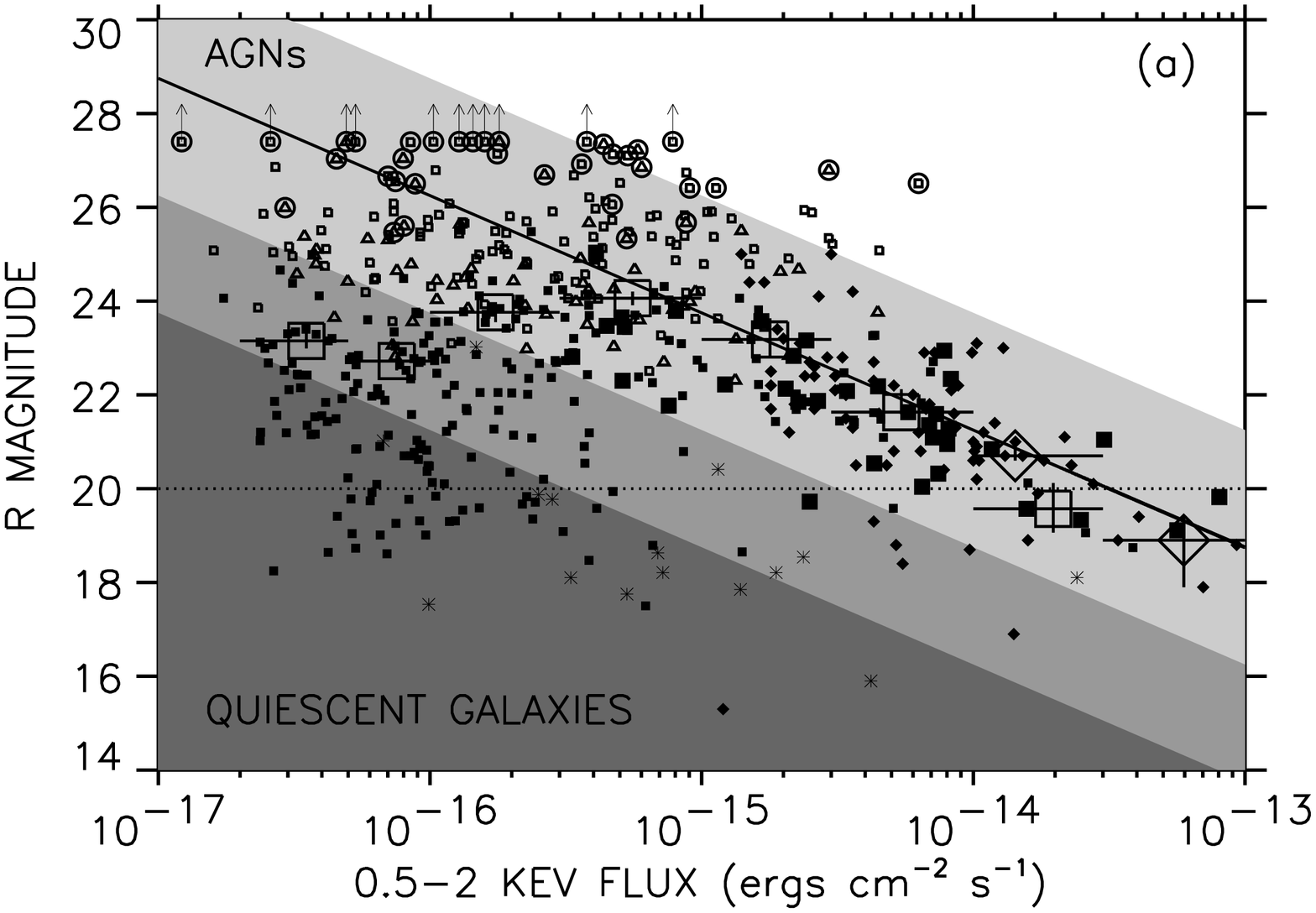,angle=90,width=5.3in}}
\centerline{\psfig{figure=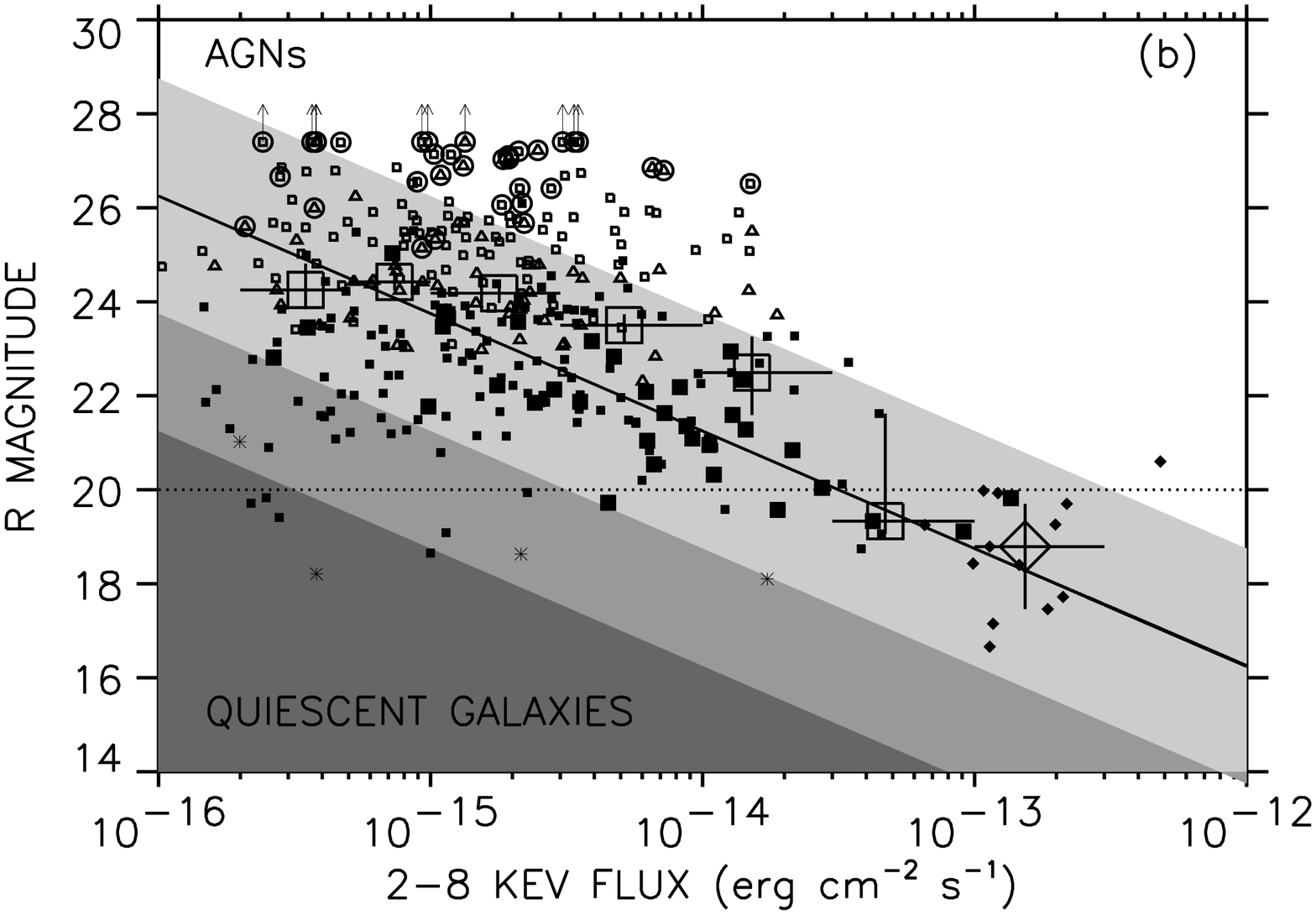,angle=90,width=5.3in}}
\vspace{6pt}
\figurenum{2}
\caption{
$R$ magnitude vs. $0.5-2$~keV flux for the CDF-N soft X-ray sources
({\it solid squares}---spectroscopic redshifts;
{\it open triangles}---photometric redshifts;
{\it open squares}---no redshift information;
{\it large solid squares}---broad-line AGNs;
{\it circles}---EROs with $R-HK'>5.7$;
{\it asterisks}---stars)
and the Lehmann et al.\ (2001) {\it ROSAT} data
({\it solid diamonds}). Large open squares
(large open diamonds) with uncertainties show the median values
of the optical magnitudes for the CDF-N (Lehmann et al.) X-ray sources.
(b) $R$ magnitude vs. $2-8$~keV flux for the CDF-N hard
X-ray sources ({\it symbols as in [a]})
and the Akiyama et al.\ (2000) {\it ASCA} data
({\it solid diamonds}). Large open squares
(large open diamond) with uncertainties show the median values
of the optical magnitudes for the CDF-N
(Akiyama et al.) X-ray sources.
In both (a) and (b), the solid line is $\log (f_X/f_R)=0$.
Shaded regions indicate typical
values of X-ray-to-optical flux ratios for different source classes:
AGNs typically lie in the lightest shaded region
[$\log (f_X/f_R)=\pm 1$]
and quiescent galaxies in the darkest shaded region
[$\log (f_X/f_R)<-2$]. Between these is a transitional
zone populated by starburst galaxies and AGNs ({\it medium shading}).
Magnitudes brighter than $R\sim 20$ ({\it dotted line}) in the CDF-N
sample suffer from saturation problems and are likely to be
underestimated.
\label{fig2}
}
\end{figure*}

In Figure~\ref{fig2}a we plot $R$ magnitude versus $0.5-2$~keV 
flux for the CDF-N soft X-ray sources. 
At bright X-ray fluxes, we also plot the 
\markcite{lehmann01}Lehmann et al.\ (2001) {\it ROSAT}
Ultra Deep Survey data ({\it solid diamonds}; groups
and clusters and one optically unidentified source have
been excluded). Spectroscopically identified stars are denoted 
by asterisks. We note in passing that flares of at least factors 
of several are observed in some of the stars
(E. D. Feigelson et al., in preparation),
as can be seen from their high $f_{0.5-2~{\rm keV}}/f_R$ ratios. 

In Figure~\ref{fig2}b we plot $R$ magnitude versus $2-8$~keV
flux for the CDF-N hard X-ray sources. At bright X-ray fluxes,
we include the \markcite{akiyama00}Akiyama et al.\ (2000)
{\it ASCA} Large Sky Survey data ({\it filled diamonds}; the
two clusters and one source without an optical identification
have been excluded, and the star is off-scale), after converting
their $2-10$~keV fluxes to $2-8$~keV assuming $\Gamma=1.7$
(the value Akiyama et al.\ assumed in their paper for the
intrinsic photon index).

The shaded regions in Figure~\ref{fig2}
indicate typical X-ray-to-optical flux
ratio ranges for different source classes.
X-ray-to-optical flux ratios were found to be a good way
to discriminate between source classes at bright
X-ray fluxes 
($f_{0.3-3.5~{\rm keV}}\ge 10^{-13}$~ergs~cm$^{-2}$~s$^{-1}$)
from the {\it Einstein Observatory} Extended
Medium-Sensitivity Survey data
(e.g., \markcite{maccacaro88}Maccacaro et al.\ 1988);
their discriminating power was later confirmed at greater
depths using ultradeep {\it ROSAT}
(\markcite{schmidt98}Schmidt et al.\ 1998) and, finally,
{\it Chandra} data (\markcite{horn01}Hornschemeier et al.\ 2001;
\markcite{barger02}Barger et al.\ 2002).
The flux in the $R$ band, $f_R$, is related to the $R$
magnitude by $\log f_R = -5.5 - 0.4 R$.
AGNs typically lie in the regions defined by the loci
$\log(f_X/f_R)=\pm 1$ ({\it lightest shading}). 
We see from Figure~\ref{fig2}
that this trend continues to hold for a large number of
sources down to very faint optical magnitudes and X-ray fluxes.

Median optical magnitudes for the CDF-N X-ray sources
({\it large open squares}) and for the
\markcite{lehmann01}Lehmann et al.\ (2001) (Fig.~\ref{fig2}a) 
or \markcite{akiyama00}Akiyama et al.\ (2000) (Fig.~\ref{fig2}b)
sources ({\it large open diamonds}) are also shown.
The horizontal bars show
the widths of the flux bins, while the vertical bars show
the 68\% confidence ranges in the medians computed using the
number of sources in each bin
(\markcite{gehrels86}Gehrels 1986). 

In Figure~\ref{fig2}a, at bright X-ray fluxes,
$\log(f_{{0.5-2}~{\rm keV}}/f_R)=0$ matches the median optical 
magnitudes because the total light output from the sources is 
dominated by unobscured AGNs. 
At $f_{{0.5-2}~{\rm keV}}\lesssim 3\times 10^{-16}$~ergs~cm$^{-2}$~s$^{-1}$,
quite a few sources whose optical counterparts are brighter
than expected for AGNs begin to populate the sample. The median
optical magnitudes drop as host galaxy light from this growing
population of normal galaxies begins to dominate the total
light output and the ratio of the X-ray to optical light decreases.
By $\log(f_{{0.5-2}~{\rm keV}}/f_R)<-2$ ({\it darkest shading}),
most of the sources are fairly normal galaxies whose X-ray
emission is dominated by processes associated with star
formation and accreting binary systems
(\markcite{horn03}Hornschemeier et al.\ 2003).

In Figure~\ref{fig2}b, at the brightest X-ray fluxes,
$\log(f_{2-8~{\rm keV}}/f_R)=0$ also matches the median optical
magnitudes. However, at fainter X-ray fluxes, the medians deviate
above $\log(f_{2-8~{\rm keV}}/f_R)=0$ as the sources become obscured
in the optical due to dust and gas. At the faintest X-ray fluxes,
the medians flatten as the optical light from the host galaxy
begins to dominate the total light output from each source.

\section{Spectroscopic Redshifts}
\label{secz}

The spectroscopic observations described in
\markcite{barger02}Barger et al.\ (2002) were made with the
Low-Resolution Imaging Spectrograph (LRIS;
\markcite{oke95}Oke et al.\ 1995) on the Keck
10~m telescopes and with the HYDRA
spectrograph (\markcite{barden94}Barden et al.\ 1994) on
the WIYN\footnote{The WIYN Observatory is a joint facility
of the University of Wisconsin, Indiana University, Yale University,
and the National Optical Astronomy Observatory.} 3.5~m telescope.
In this paper we present new spectroscopic observations that
were obtained with the Deep Extragalactic Imaging Multi-Object
Spectrograph (DEIMOS; \markcite{faber02}Faber et al. 2002)
on Keck~II the nights of UT 2003 January 29--30, March 27, and
April 25--27. The observations
were made with the 600 lines per mm grating, giving a resolution
of $3.5$~\AA\ and a wavelength coverage of $5300$~\AA. The spectra
were centered at an average wavelength of $7200$~\AA, although the
exact wavelength range for each spectrum depends on the slit position.
Each $\sim 1$~hr exposure was broken
into three subsets, with the objects stepped along the slit by
$1.5''$ in each direction. The spectra were reduced in the same
way as previous LRIS spectra
(\markcite{cowie96}Cowie et al.\ 1996).

%
%

\newpage

In the vast majority of cases, only spectra that could be
confidently identified based on multiple emission and/or absorption
lines were included. However, we have assigned redshifts to thirteen
sources based primarily on a single line and the
continuum shape. These sources may be less reliable than the other
sources in the catalog. We have noted these sources in Table~A1.

We have only cross-identified X-ray sources with spectroscopic
counterparts if the radial offsets are $\le 2''$.
The only exceptions are the seven off-axis X-ray sources
discussed in \S~\ref{secimaging}. We have placed the optical
positions of these off-axis sources
(sources 121, 197, 270, 316, 404, 410, and 479)
at the bright galaxy nuclei.
[We note that the redshift for object 316 was given
incorrectly in \markcite{barger02}Barger et al.\ (2002; their
object 218) to be $z=0.230$; the correct redshift is $z=0.213$.]

%
%
\begin{inlinefigure}
\psfig{figure=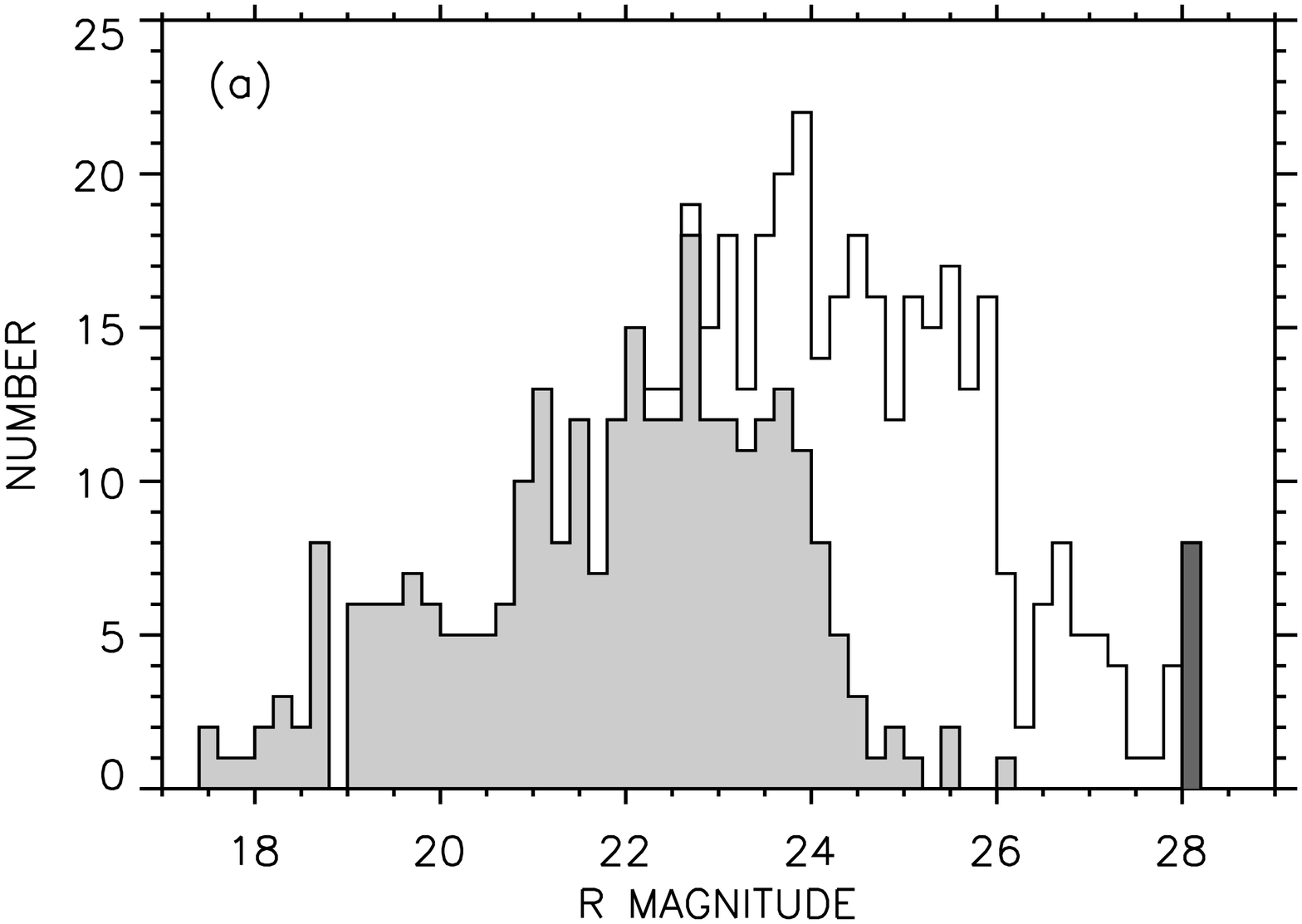,angle=90,width=3.5in}
\psfig{figure=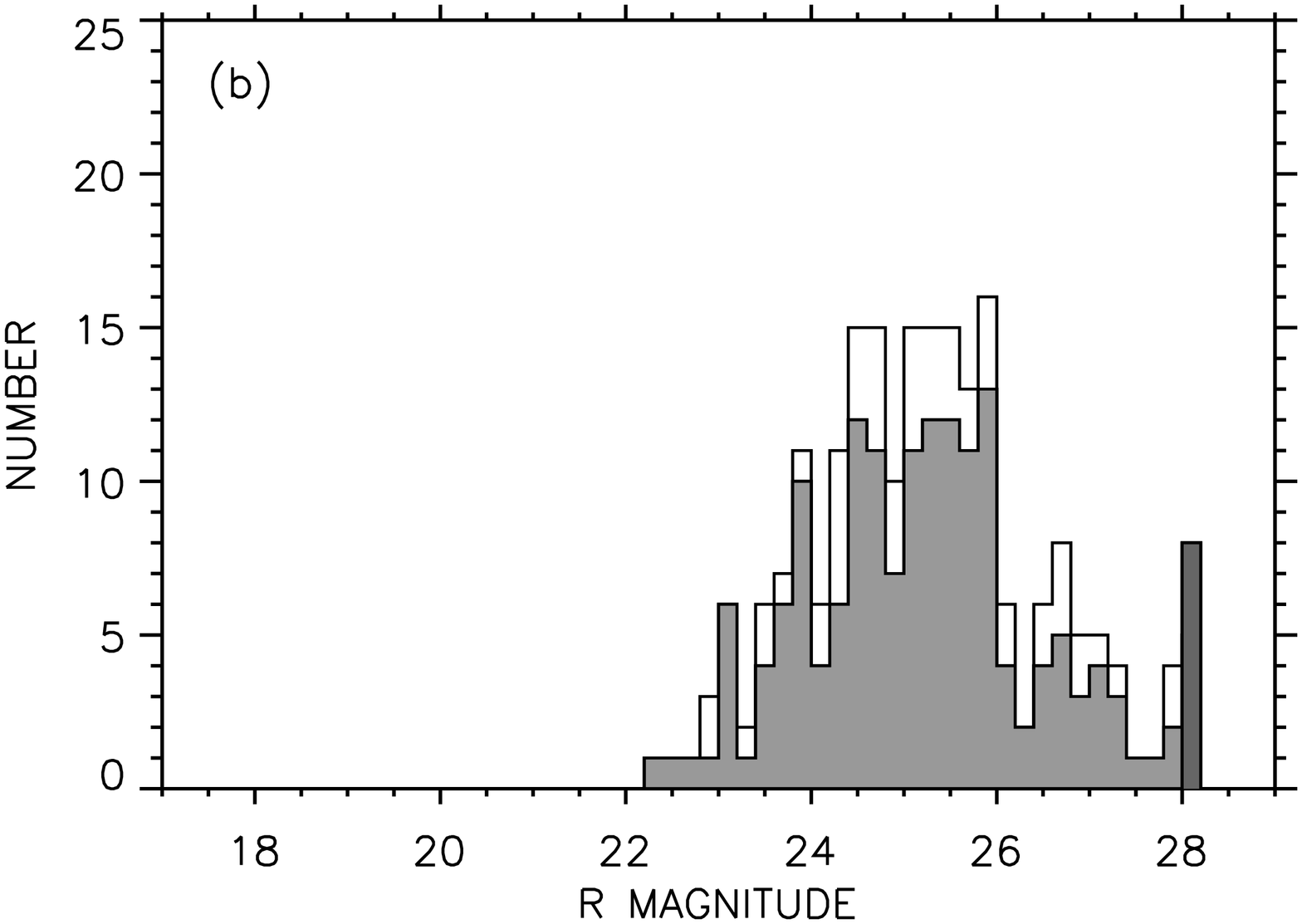,angle=90,width=3.5in}
\vspace{6pt}
\figurenum{3}
\caption{
(a) $R$ magnitude distribution for the 503 CDF-N X-ray 
sources ({\it open}). Light shaded region denotes the 284 sources 
with spectroscopic identifications. Dark shaded region denotes 
sources fainter than the $1\sigma$ image limit of $R=28.1$.
Magnitudes brighter than $R\sim 20$ suffer from saturation 
problems and are likely to be underestimated.
(b) $R$ magnitude distribution for the 219 CDF-N X-ray
sources that either were not observed spectroscopically 
or were observed but could not be identified ({\it open}).
Medium shaded region denotes the 164 sources that were
observed but could not be identified. Dark shaded region is 
as in (a); five of these sources were observed.
\label{fig3}
}
\addtolength{\baselineskip}{10pt}
\end{inlinefigure}

With this selection, 284 of the 503 X-ray sources (56\%)
have spectroscopic identifications, including 14 stars.
Thirty-six of the sources have broad emission lines.
Three of the redshifts (sources 173, 219, and 378) are taken 
from the compilation of \markcite{cohen00}Cohen et al.\ (2000).
In Figure~\ref{fig3}a we show the $R$ magnitude distribution 
for all 503 sources ({\it open}). We 
indicate which of the sources have spectroscopic identifications 
with the light shaded region.
Of the 299 sources with $R\le 24$, 261 (87\%)
have been spectroscopically identified, and the bulk of
those that have not been identified lie in the magnitude range
$R=23-24$. In Figure~\ref{fig3}b we show the $R$ magnitude 
distribution for the 219 CDF-N X-ray sources that either were 
not observed spectroscopically ({\it open}) or were observed 
but could not be identified ({\it medium shading}). 

\section{Photometric Redshifts}
\label{secphotz}

Broadband galaxy colors have been used in recent years to
obtain photometric redshift estimates for galaxies
(see \markcite{bol00}Bolzonella, Miralles, \& Pell{\'o} 2000 
for a review). X-ray sources may have complex spectral energy 
distributions that arise from both the host galaxy and 
the AGN. Consequently, estimating photometric redshifts for 
X-ray sources using standard galaxy templates may be difficult. 
However, we are most interested in obtaining photometric
redshifts for obscured AGNs whose spectroscopic signatures 
are not easily visible, and for those sources, photometric 
redshift estimates based on standard galaxy templates may be 
expected to work well.

\markcite{capak03b}Capak et al.\ (2003b) estimated photometric
redshifts for sources in the CDF-N region using 
the template fitting method and photometry in the
$U$, $B$, $V$, $R$, $I$, $z'$, and $HK'$ bands. 
This method should avoid the 
biases at high redshifts that are introduced by the training 
set method (\markcite{connolly}Connolly et al.\ 1995)
or the principal component analysis method
(\markcite{cabanac}Cabanac, de Lapparent, \& Hickson 2002).
Capak et al.\ used their large spectroscopic database to 
compare results from two publicly available
codes, HYPERZ (\markcite{bol00}Bolzonella et al.\ 2000) and
BPZ (Bayesian photometric redshift estimation;
\markcite{benitez}Ben{\'i}tez 2000).
The major difference between the codes is the inclusion of a 
weighting function in BPZ that reduces the number of degenerate 
fits at different redshifts. Additional differences are that 
HYPERZ includes a range of templates from the updated
Bruzual \& Charlot evolutionary code (GISSEL98;
\markcite{bc93}Bruzual \& Charlot 1993), as well as the 
\markcite{cww}Coleman, Wu, \& Weedman (1980; hereafter, CWW) 
templates, while BPZ only uses the templates of CWW and 
\markcite{kinney}Kinney et al.\ (1996); however, BPZ 
does interpolate between the templates to produce 
intermediate templates. Also, an intrinsic reddening term 
is included in HYPERZ but not in BPZ, and the two codes 
treat non-detections differently.
For most of the galaxies in the $0<z<1$ range, 
\markcite{capak03b}Capak et al.\ (2003b)
found that the two codes produced similar results; however, 
HYPERZ confused a larger number of sources at $z<0.5$ with 
$z>3$ sources, which is not surprising since the weighting 
function in BPZ was introduced to solve this problem.

\markcite{capak03b}Capak et al.\ (2003b) were able to improve 
the results of both codes by tuning the photometry. They 
fitted templates at the known redshift for each source and
then calculated the mean offset between their photometry 
and the template photometry. Iterating on this process produced 
small photometric offsets (0.049 in $U$, 0.022 in $V$,
0.069 in $R$, $-0.064$ in $z'$, and $-0.196$ in $HK'$)
that significantly improved the results
(see Capak et al.\ 2003b for details);
the $B$ and $I$ band magnitudes were fixed since the authors 
were confident in the photometry in those bands. The offsets
are likely due to differences between real galaxies and the 
templates and between the real instrument filter profiles and 
the assumed profiles.

The BPZ code outputs the probability $p_{\Delta z}$
of $|z-z_b|<\Delta z$, where $z$ is the galaxy redshift 
and $z_b$ is the ``best'' redshift estimate. When the
value of $p_{\Delta z}$ is low, the redshift probability 
is spread over a large range in redshift, and
the prediction is likely not reliable, so we applied a
$p_{\Delta z}>0.90$ threshold to remove galaxies with
catastrophic redshift errors (see 
\markcite{capak03b}Capak et al.\ 2003b for more details).
We also applied a $R>20.5$ threshold to eliminate saturated 
sources. This is slightly fainter than the saturation limit 
in $R$ to also remove sources that may only be saturated in other 
bands. Such sources would not have good photometric redshift 
estimates, but since they are already spectroscopically 
identified, photometric redshift estimates are not needed.

%
%
\begin{inlinefigure}
\psfig{figure=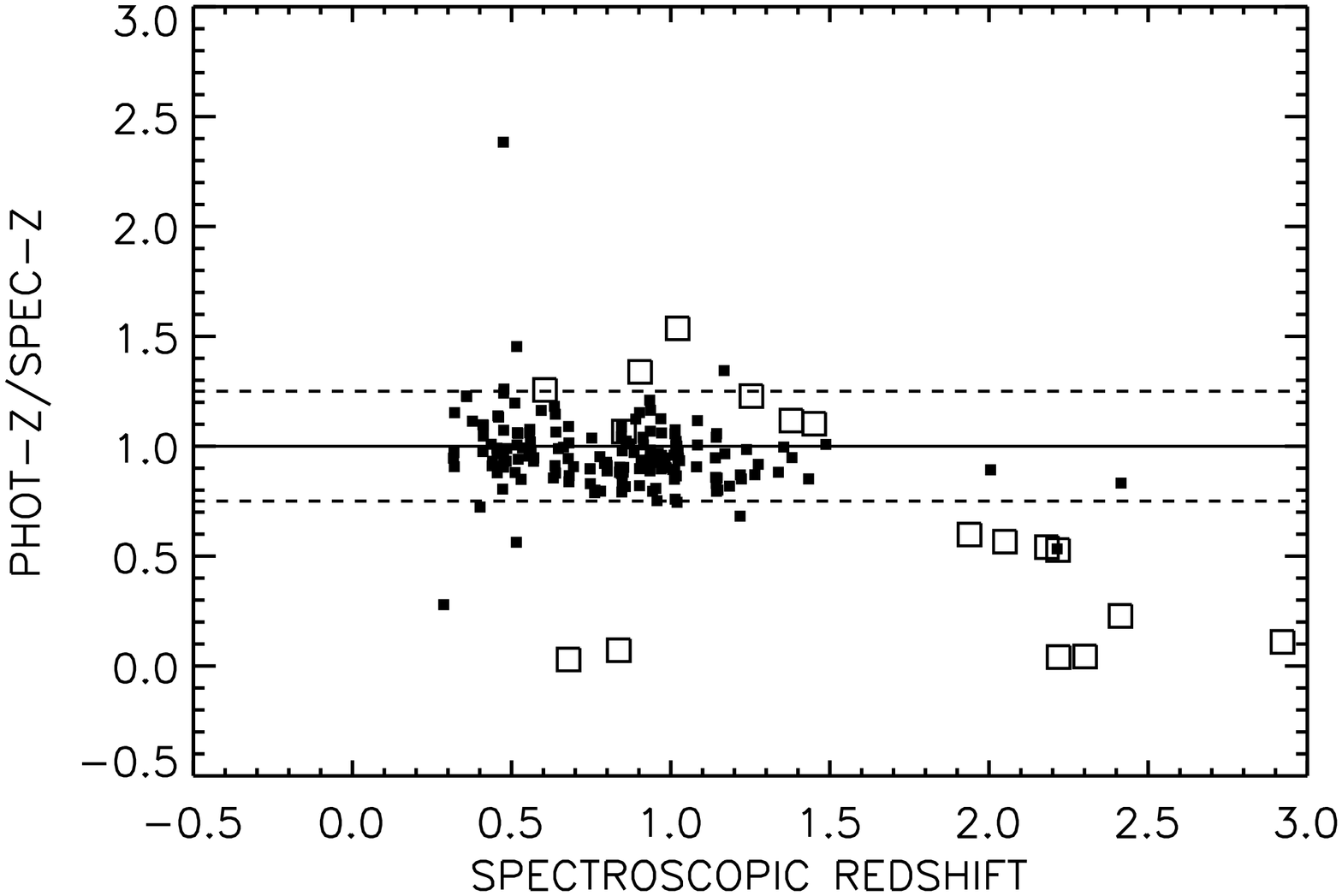,angle=90,width=3.5in}
\vspace{6pt}
\figurenum{4}
\caption{
Ratio of photometric (calculated with the BPZ code) to
spectroscopic redshift vs. spectroscopic redshift for
the CDF-N X-ray sources with both measurements, excluding saturated
sources (63) and sources with BPZ probabilities $<0.90$ (28). 
Only one source (source 173 at $z=0.08$) is not within the vertical 
range of the plot. Broad-line AGNs ({\it open squares}) are not expected 
to have good photometric redshift estimates from galaxy templates
due to the dominance of the unobscured AGN to the total light output. 
Photometric redshift errors are accentuated for sources at low
redshifts due to the ratioing, but
the photometric redshifts agree with the spectroscopic redshifts
to within 25\% for 94\% of the non-broad-line sources
with both measurements.
\label{fig4}
}
\addtolength{\baselineskip}{10pt}
\end{inlinefigure}

In Figure~\ref{fig4} we plot the ratio of the photometric
redshift estimate from BPZ to the spectroscopic
redshift versus the spectroscopic redshift for the 193 remaining
X-ray sources (172 non-broad-line and 21 broad-line sources)
with both measurements. We plot the redshift ratio
because what we are interested in is how accurate our distances
are. This means that photometric redshift errors at low redshifts
are accentuated. Some of the discrepant objects at low redshifts
have complex optical structures near
the X-ray source positions that contaminate the photometry.
The broad-line AGNs with both spectroscopic and photometric
redshifts are denoted by open squares. The photometric redshift
method fails for many of them, but broad-line AGNs are
straightforward to identify spectroscopically, even in
the redshift range $z\sim 1.5-2$, and we now have observed 
the bulk of the sources in the CDF-N X-ray sample 
(see Figure~\ref{fig3}b), so we do not need to
worry that broad-line AGNs will contaminate our photometric
redshift sample.

The photometric redshifts for the non-broad-line sources are
clearly quite robust, with 94\% having photometric 
redshifts within 25\% ({\it horizontal dashed lines})
of their spectroscopic redshifts. Only one very low redshift 
source has a badly discrepant photometric redshift 
(spectroscopic redshift of $z=0.08$ and photometric redshift 
of 2.58) and hence is not within the vertical range of the plot.
We have not attempted to refine the photometric redshifts 
with spectral information (e.g., there may be cases where we 
had insufficient spectral information to reliably identify
the source spectroscopically but that information 
could have been used to try to improve the photometric redshifts), 
since for most purposes the photometric redshifts are accurate 
enough. The BPZ photometric redshifts presented in this paper 
are an improvement over the photometric redshift estimates 
made with HYPERZ for the 1~Ms CDF-N X-ray sources
(\markcite{barger02}Barger et al.\ 2002).

\section{Redshift Distribution}
\label{seczdist}

%
%
\begin{inlinefigure}
\psfig{figure=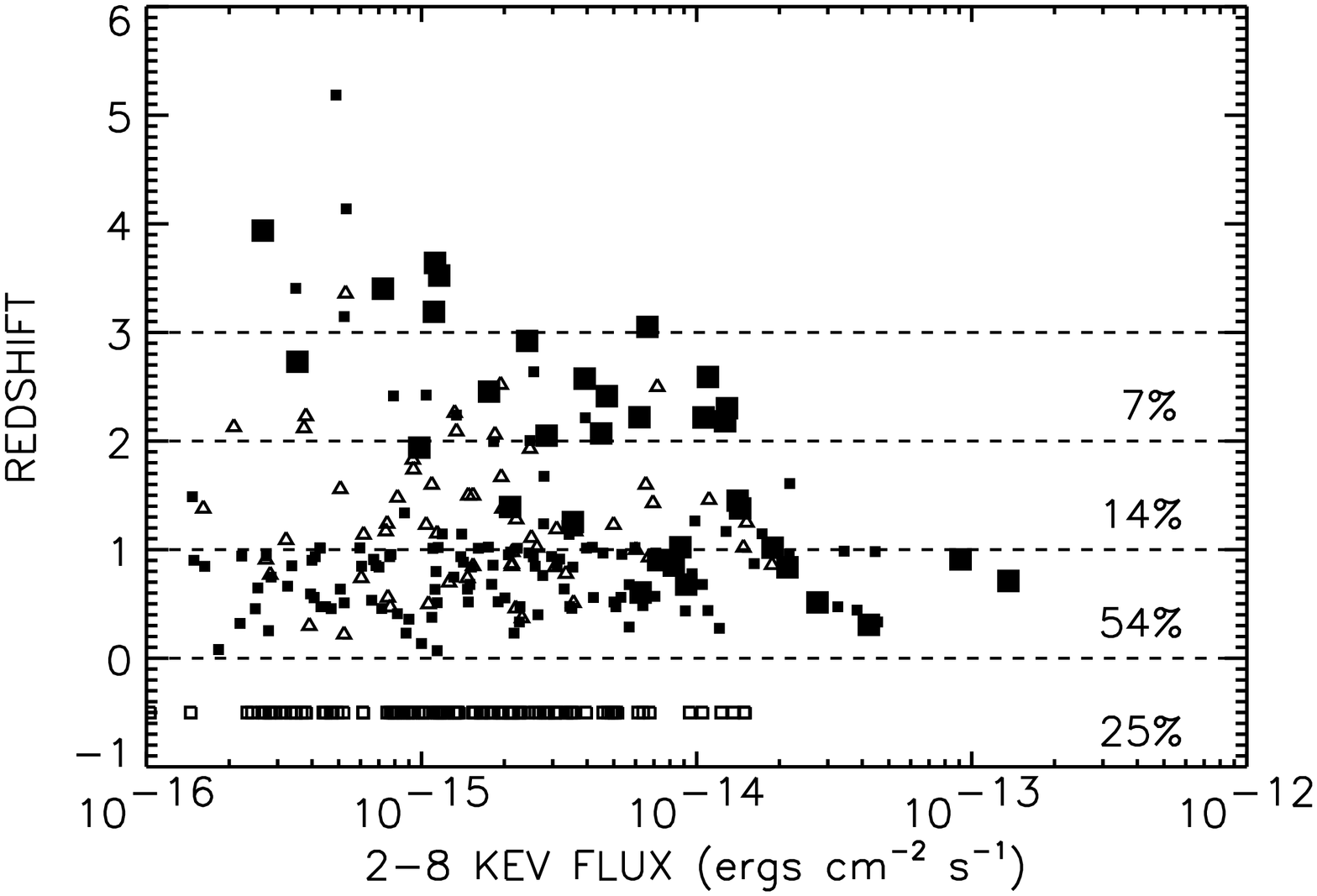,angle=90,width=3.5in}
\vspace{6pt}
\figurenum{5}
\caption{
Redshift vs. $2-8$~keV flux for the CDF-N hard X-ray
sources. Sources with spectroscopic ({\it solid squares}) or
photometric ({\it open triangles}) redshifts are plotted at
those redshifts, while those without are plotted below
the $z=0$ line ({\it open squares}). Broad-line AGNs
are denoted by large solid squares. Percentages
of the measured hard X-ray light from the combined deep and
bright subsamples that come from only spectroscopically
identified sources in each redshift interval are given
at the right.
\label{fig5}
}
\addtolength{\baselineskip}{10pt}
\end{inlinefigure}

In Figure~\ref{fig5} we show redshift versus hard X-ray flux
for the spectroscopically ({\it solid squares}) and photometrically
({\it open triangles}) identified CDF-N hard X-ray sources. 
To avoid incompleteness at the faint end when determining the
fractional light contribution from each redshift interval to
the measured hard X-ray light, we consider two restricted
uniform flux-limited hard X-ray subsamples, which we refer
to as our ``bright'' and ``deep'' subsamples.
For the bright subsample we consider a
$10'$ radius region around the approximate center of the
X-ray image and select sources detected with
fluxes above $3\times 10^{-15}$~ergs~cm$^{-2}$~s$^{-1}$.
[We note, however, that above $10^{-13}$~ergs~cm$^{-2}$~s$^{-1}$ 
the number density of sources in the CDF-N becomes too low for 
an accurate measurement.] For the deep subsample we consider 
a $6'$ radius high image quality and high exposure time region 
and select sources detected with fluxes between $3\times 10^{-16}$ 
and $3\times 10^{-15}$~ergs~cm$^{-2}$~s$^{-1}$.

The numbers at the right of Figure~\ref{fig5} are the percentages
of the measured hard X-ray light from the combined deep and bright
subsamples that come from only spectroscopically identified sources
in each redshift interval. The spectroscopically identified sources
already comprise 75\% of the measured hard X-ray light, of which
at least 54\% arises at redshifts below $z=1$ and at least
68\% below $z=2$. These percentages increase to 58\% and 76\%,
respectively, if we include the sources with photometric redshifts,
and they rise even higher if the bright X-ray sources found in
deep {\it ASCA} surveys (\markcite{akiyama00}Akiyama et al.\ 2000)
are considered (see Figure~14a of \markcite{barger02}Barger et al.\ 2002).
Thus, an impressive amount of the $2-8$~keV extragalactic
background light arises at recent times, as was first shown
by \markcite{barger01a}Barger et al.\ (2001a).

In Figure~\ref{fig6} we show the redshift-magnitude relation for
the CDF-N X-ray sources; sources
without any redshift information are plotted below $z=0$.
Broad-line AGNs ({\it large solid squares}) are systematically
the most optically luminous of the X-ray sources because of their
AGN contribution to the light. The sources with photometric
redshifts show a smooth continuation of the spectroscopic
trend towards fainter $R$ magnitudes with increasing redshift.

%
%
\begin{inlinefigure}
\psfig{figure=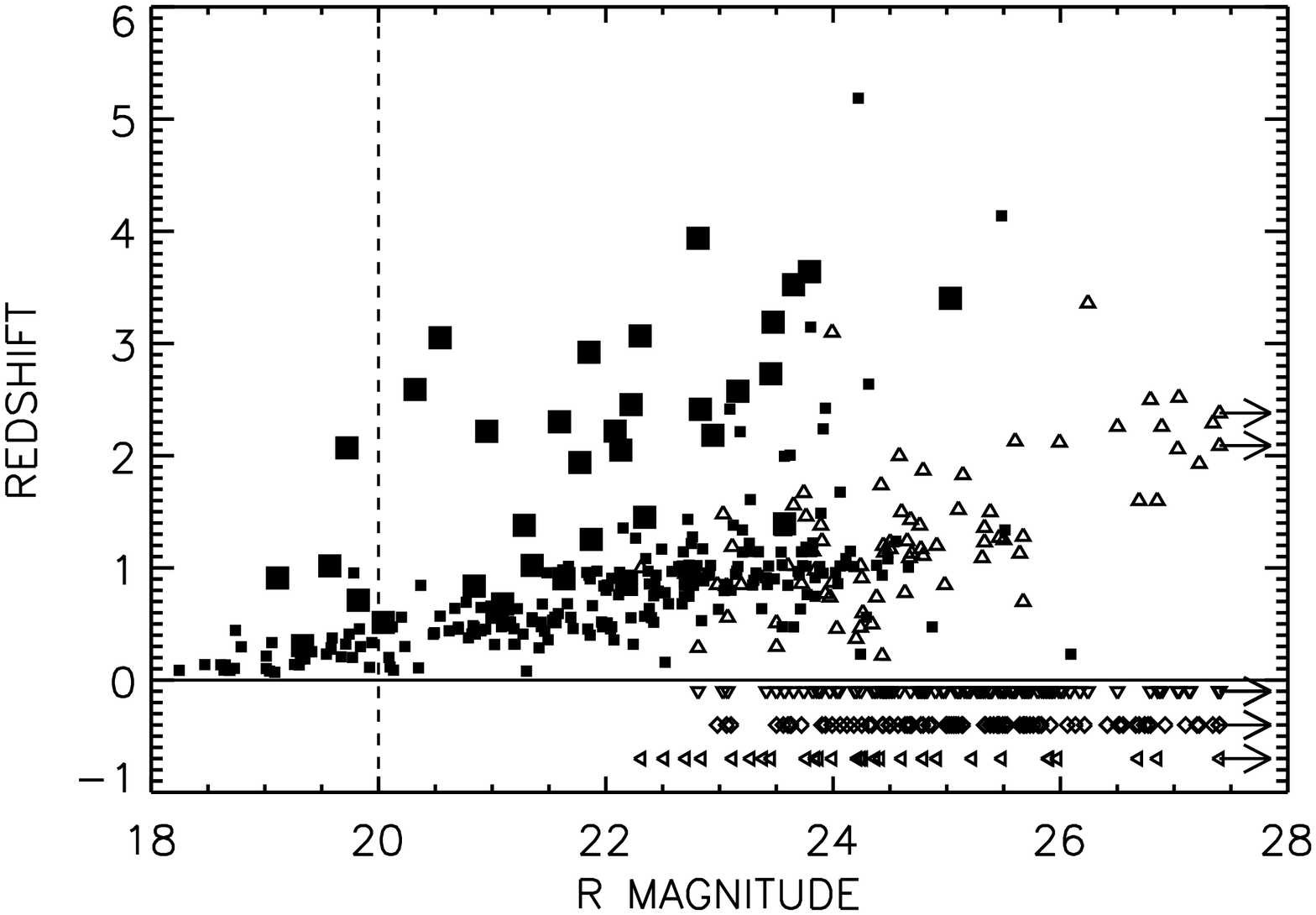,angle=90,width=3.5in}
\vspace{6pt}
\figurenum{6}
\caption{
Redshift vs. $R$ magnitude for the CDF-N X-ray sources, excluding
stars ({\it solid squares}---spectroscopic redshifts;
{\it open triangles}---photometric redshifts;
{\it downward pointing triangles at z=$-$0.1}---sources
without any redshift information at radii less than $6'$;
{\it diamonds at z=$-$0.4}---sources without
any redshift information at radii between $6'$ and $10'$;
{\it sideways pointing triangles at z=$-$0.7}---sources without
any redshift information at radii greater than $10'$;
{\it large solid squares}---broad-line AGNs).
Magnitudes brighter than $R\sim 20$ suffer from saturation
problems and are likely to be underestimated. Sources undetected
at the $2\sigma$ limit of $R=27.4$ are plotted at this
magnitude with rightward pointing arrows.
\label{fig6}
}
\addtolength{\baselineskip}{10pt}
\end{inlinefigure}

%
%
\begin{inlinefigure}
\psfig{figure=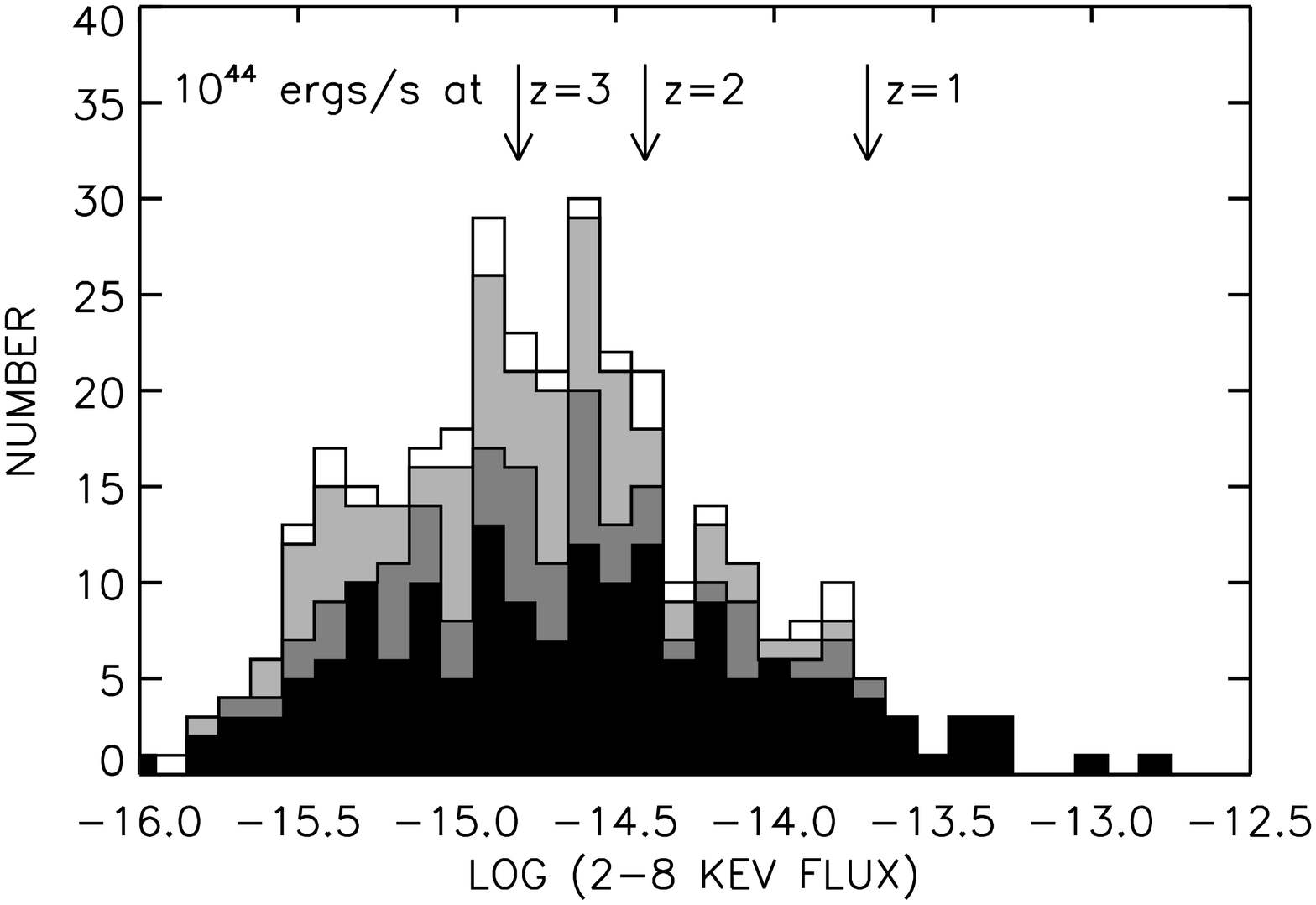,angle=90,width=3.5in}
\vspace{6pt}
\figurenum{7}
\caption{
$2-8$~keV flux distribution for the 503 CDF-N X-ray sources
({\it black shading}---spectroscopic redshifts;
{\it medium shading}---photometric redshifts;
{\it light shading}---spectroscopically observed but
unidentified; {\it no shading}---unobserved).
Arrows mark the flux of a source with a rest-frame $2-8$~keV 
luminosity of $10^{44}$~ergs~s$^{-1}$ at a redshift of $z=1$, 2, 
or 3, computed with a $K$-correction for an intrinsic $\Gamma=1.8$ 
power-law spectrum.
\label{fig7}
}
\addtolength{\baselineskip}{10pt}
\end{inlinefigure}

In Figure~\ref{fig7} we show the $2-8$~keV flux distribution
for the CDF-N X-ray sources. Labeled arrows
at the top of the figure show the flux of a source with
a rest-frame $2-8$~keV luminosity of $10^{44}$~ergs~s$^{-1}$
at a redshift of $z=1$, 2, or 3, computed using the $2-8$~keV
flux and a $K$-correction for an intrinsic $\Gamma=1.8$
power-law spectrum, where $\Gamma$ is the photon index (see
\markcite{barger02}Barger et al.\ 2002 for the reason behind
this choice of $\Gamma$). The arrow at $z=3$ also roughly
corresponds to the $2-8$~keV flux limit of 
$2\times 10^{-15}$~ergs~cm$^{-2}$~s$^{-1}$
where the fitted X-ray number
counts give $\sim 80$\% of the $2-8$~keV light
(\markcite{cowie02}Cowie et al.\ 2002). 

Above the flux limit where roughly 80\% of the light arises, 
there are 150 sources.
Of these, the 113 with spectroscopic or photometric identifications
account for most of the light; only 15\% of the light 
is not accounted for by the identified sources. Sources with 
rest-frame $2-8$~keV luminosities greater than 
$10^{44}$~ergs~s$^{-1}$ (often called quasars) contribute 
39\% of the light, while fainter sources contribute 45\%. 
Of the 25 sources in the quasar luminosity range, 
15 are broad-line AGNs, and these 15 contribute 26\% 
of the light. Six of the sources in the quasar luminosity range
(sources 89, 240, 259, 390, 420, and 495)
are type~II AGNs, based on the presence of [NeIII], [NeV], or 
CIV emission lines in the spectra, though in many cases these
lines are weak features superimposed on a much brighter
galaxy spectrum. In one of these sources (source 420) the CIV 
line shows a P-Cygni profile. These six type~II quasars contribute
only 11\% of the light and hence are not the dominant
contributors to the X-ray background at these energies.
One additional source in the quasar luminosity range (source 398)
is a Ly$\alpha$ emitter with absorption at CIV, 
while the remaining three (sources 108, 165, and 463) 
have only photometric redshifts.

%
%
\begin{inlinefigure}
\psfig{figure=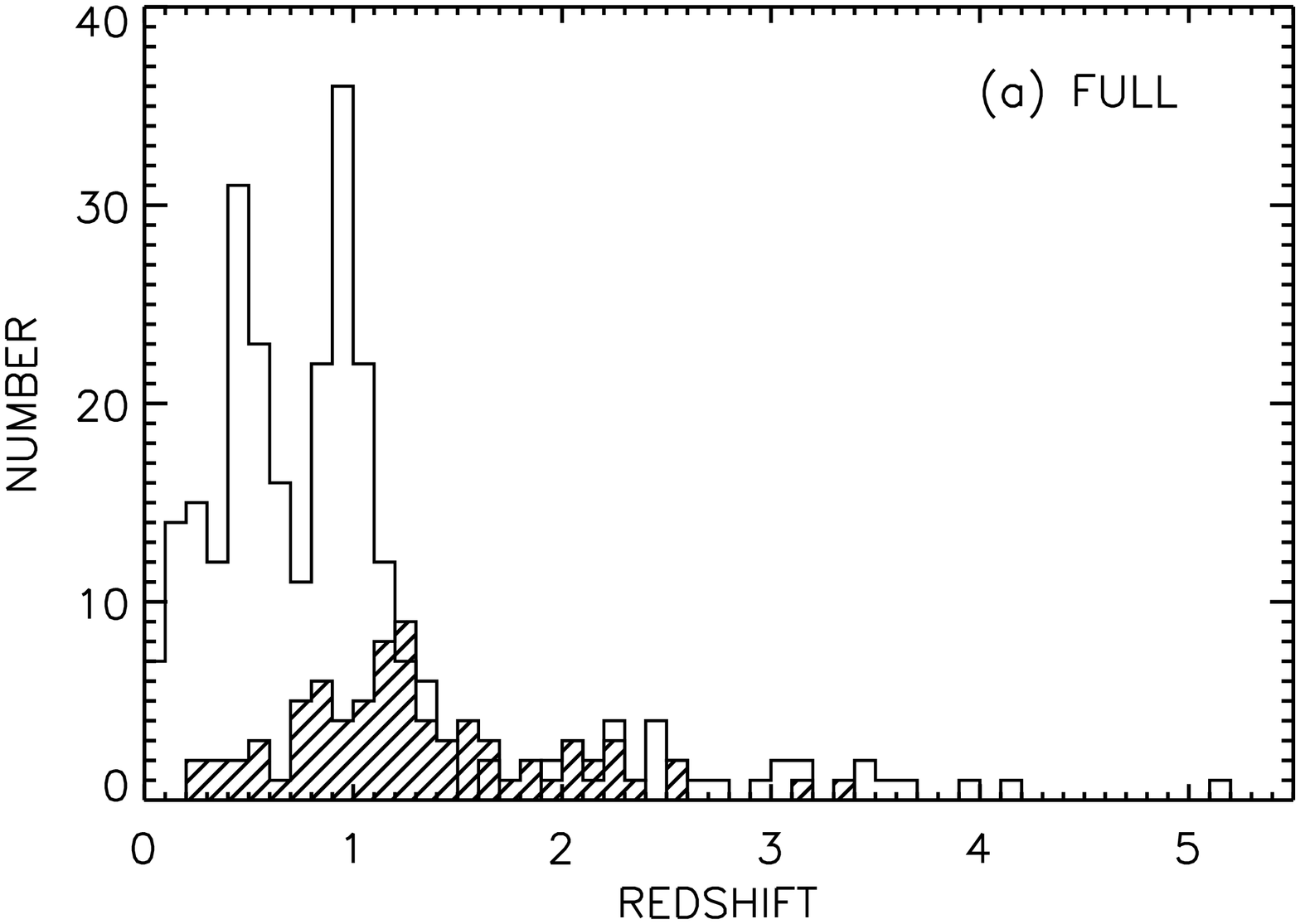,angle=90,width=3.5in}
\psfig{figure=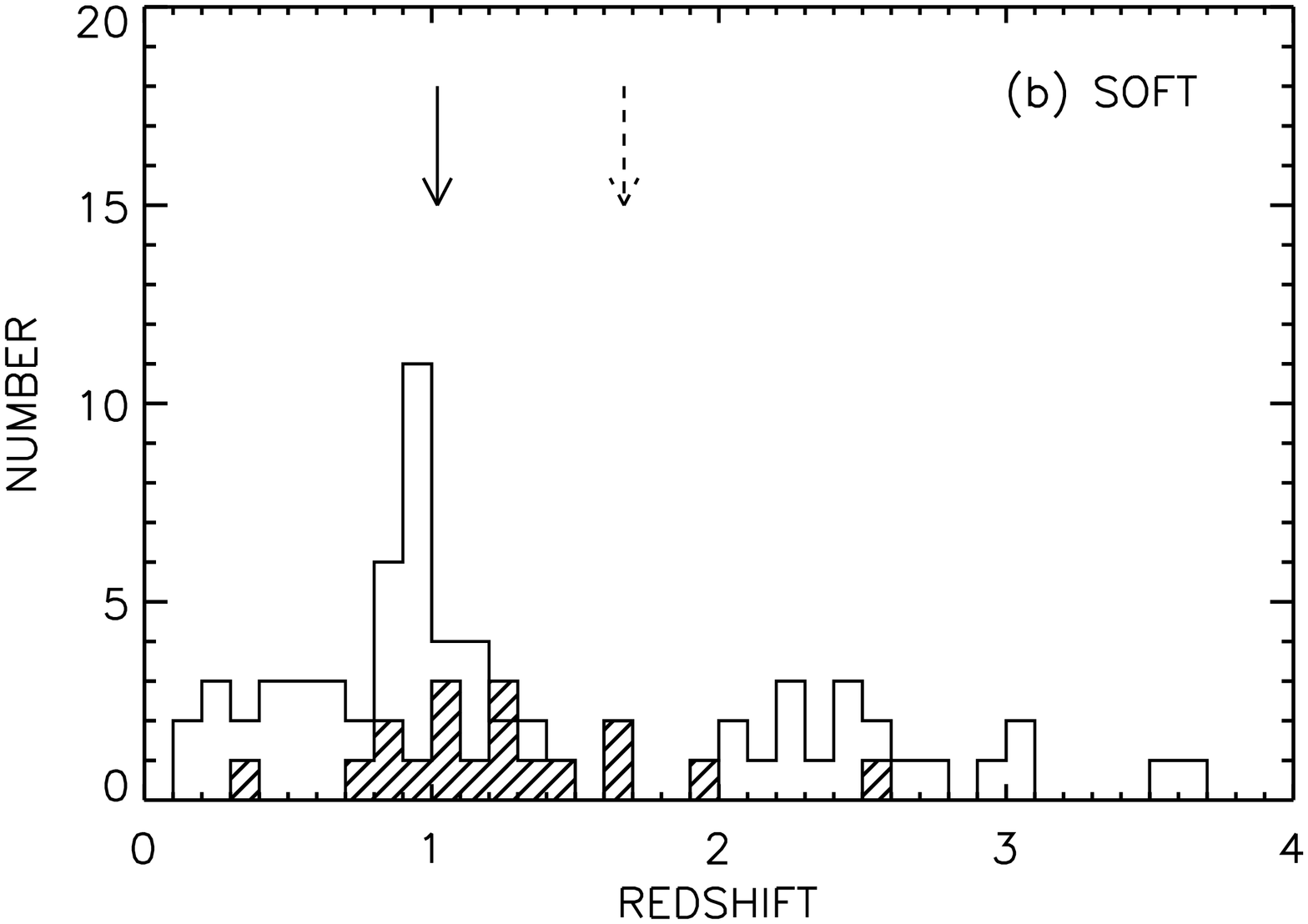,angle=90,width=3.5in}
\psfig{figure=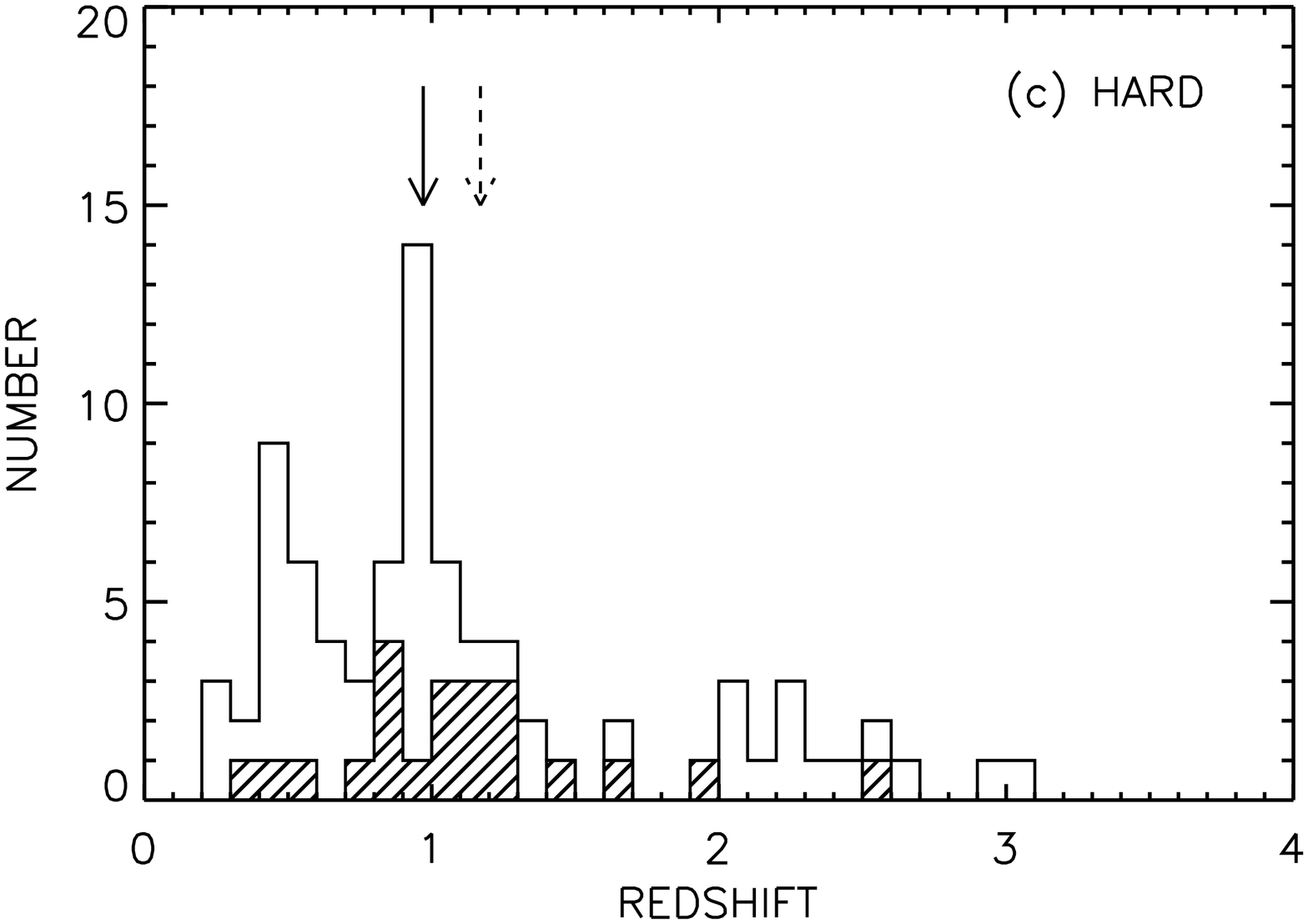,angle=90,width=3.5in}
\vspace{6pt}
\figurenum{8}
\caption{
Spectroscopic ({\it open}) and photometric ({\it hatched})
redshift distributions for (a) the CDF-N X-ray sources,
(b) the 
$f(0.5-2~{\rm keV})>5\times10^{-16}$~ergs~cm$^{-2}$~s$^{-1}$
soft X-ray sources, and 
(c) the $f(2-8~{\rm keV})>2\times 10^{-15}$~ergs~cm$^{-2}$~s$^{-1}$
hard X-ray sources. Binning is $\Delta z=0.1$.
All stars and sources without any redshift information have 
been excluded. In (b) and (c) only sources within $12'$ of the 
approximate X-ray image center are used.
Solid arrows in (b) and (c) show the median redshifts for all
the sources in the sample with spectroscopic or photometric
redshifts, and dashed arrows show the median redshifts after
placing the remaining sources without any redshift information 
at high redshifts.
\label{fig8}
}
\addtolength{\baselineskip}{10pt}
\end{inlinefigure}

In Figure~\ref{fig8}a we show the spectroscopic ({\it open}) and
photometric ({\it hatched}) redshift distributions for all the
CDF-N X-ray sources with redshift information using a low resolution
($\Delta z=0.1$) binning. The gap in the spectroscopic redshift
distribution between $z\sim1.5$ and 2 reflects the difficulty of
identifying galaxies with redshifts in this range, where
[OII]~3727~\AA\ has moved out of the optical window and
Ly$\alpha$~1216~\AA\ has not yet entered in. The two broad
redshift spikes seen in the 1~Ms exposure
(\markcite{barger02}Barger et al.\ 2002) have
grown substantially in the 2~Ms exposure. 
Fifty-four spectroscopically identified sources
lie between $z=0.4$ and $z=0.6$, with a median redshift of $z=0.48$,
and 79 lie between $z=0.8$ and $z=1.1$, with a median redshift of
$z=0.94$. Thus, these two redshift intervals contain nearly half of 
the identified galaxies. It is hard to quantify the significance
of these peaks without knowing the true redshift distribution.
If the galaxies between $z=0$ and $z=1.2$ were uniformly distributed,
then we would expect 44 galaxies in the lower redshift bin and 55
in the higher. The $z\sim0.5$ peak could simply be the maximum in
the redshift distribution, but it appears likely that the $z\sim1$ 
peak is an unique feature of the CDF-N region. 

%
%
\begin{inlinefigure}
\psfig{figure=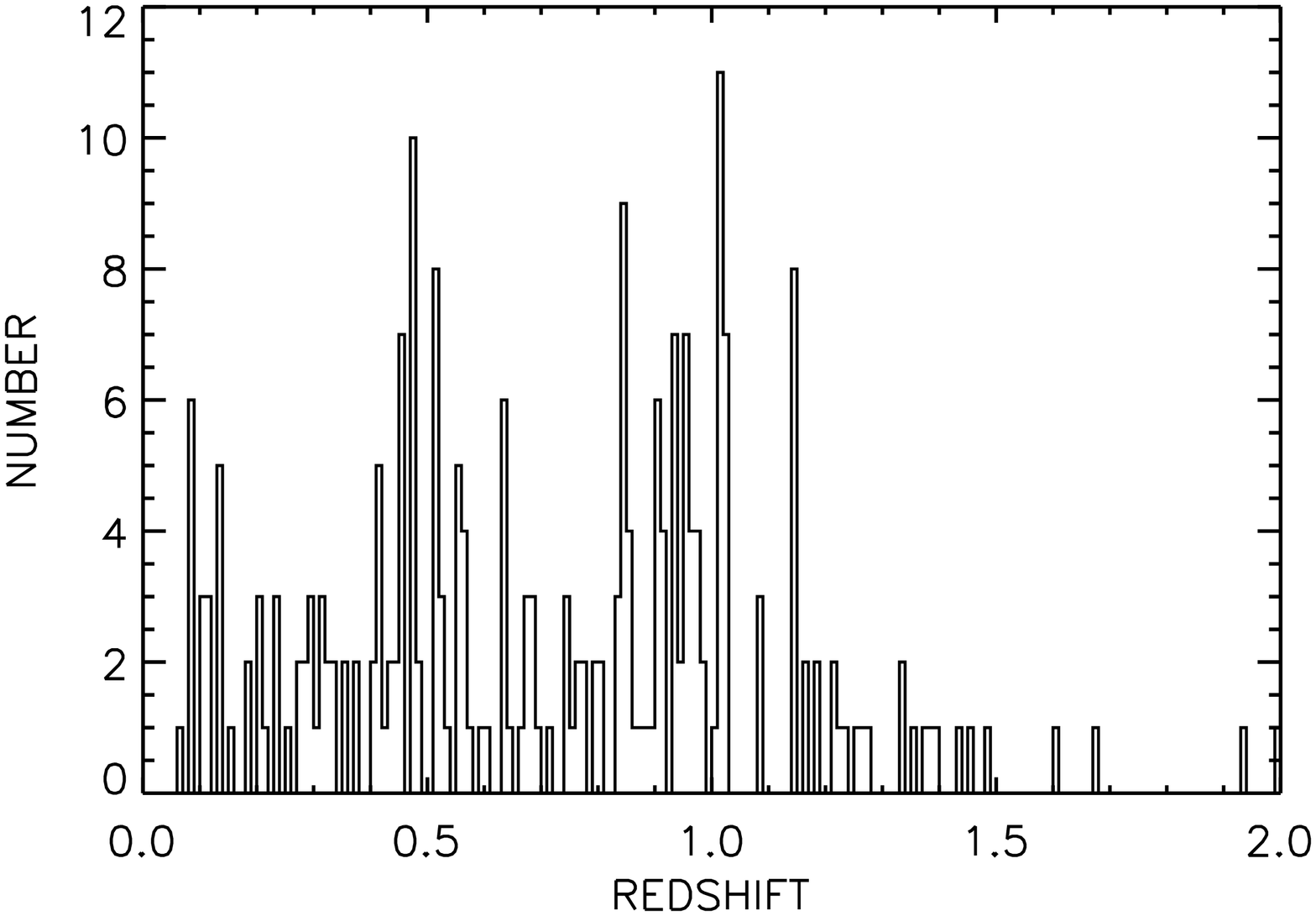,angle=90,width=3.5in}
\vspace{6pt}
\figurenum{9}
\caption{
Spectroscopic redshift distribution for the CDF-N X-ray sources
using a $\Delta z=0.01$ binning.
\label{fig9}
}
\addtolength{\baselineskip}{10pt}
\end{inlinefigure}

In Figures~\ref{fig8}b and \ref{fig8}c we show the separate 
redshift distributions for the soft and hard
samples, respectively, including only sources within $12'$ of
the approximate X-ray image center with $0.5-2$~keV fluxes above
$5\times10^{-16}$~ergs~cm$^{-2}$~s$^{-1}$ (Figure~\ref{fig8}b)
and $2-8$~keV fluxes above $2\times 10^{-15}$~ergs~cm$^{-2}$~s$^{-1}$ 
(Figure~\ref{fig8}c).
These limiting fluxes were chosen to match roughly the value
where the fitted X-ray number counts give $\sim 80$\% of the 
light (\markcite{cowie02}Cowie et al.\ 2002).
The solid arrows show the
median redshifts of all the sources with spectroscopic and
photometric redshifts in each sample, and the dashed arrows
show the median redshifts after all the sources without any
redshift information have been arbitrarily placed at high 
redshifts. The dashed arrows are only shifted 
to slightly higher redshifts than the solid arrows because most 
of the sources to these flux levels have already been identified.

In Figure~\ref{fig9} we show the redshift distribution for all
the spectroscopically identified X-ray sources between $z=0$ and 
$z=2$ using a higher resolution ($\Delta z=0.01$) binning. 
The strongest single feature in the distribution is at $z=1.0175$,
where 16 galaxies lie within a thousand km~s$^{-1}$ of the redshift.
(These are split between two bins in Fig.~\ref{fig9}
but do not contain all of the galaxies in each of the bins.)
Six of these sources lie in a tight spatial clump (source 116 is
the brightest galaxy in this clump).

Because of the apparent velocity sheets and complex structures in 
the redshift distributions of the X-ray sources in individual 
{\it Chandra} fields, we will need to observe a large number of fields 
in order to get the average true redshift distribution.

\section{Optical/Near-Infrared Colors and Magnitudes}

In Figure~\ref{fig10}a we show $B-I$ color versus $I$ magnitude
for the X-ray sources with both X-ray positional uncertainties
and optical counterpart separations $<1''$. (The tight positional
selection is used in order to keep the field contamination
low; see \S~\ref{secimaging}.) Median $B-I$ colors for the 
non-broad-line sources in each magnitude bin are shown
as large open diamonds. The 68\% confidence ranges
in the medians were computed using the number of sources in each
bin (\markcite{gehrels86}Gehrels 1986). The median $B-I$ color
for all the broad-line sources is shown as a dashed line.
The figure shows a general trend of the non-broad line sources
towards bluer colors at fainter 
magnitudes, with the colors of the $I>24$ sources approaching 
the median blue color of the broad-line AGNs.

\markcite{schreier01}Schreier et al.\ (2001), and subsequently
\markcite{koekemoer02}Koekemoer et al.\ (2002), also noted this
color trend with magnitude from {\it HST} data on the CDF-S. 
They interpreted it as meaning that there are two distinct 
populations of sources, an optically bright red population that 
consists of normal galaxies at $z<1$, and an optically faint
blue population that consists of type 2 AGNs of low to moderate 
luminosity located at $z\sim 1-2$. 
Our interpretation is instead that the optically 
faint blue population consists of normal galaxies at $z>1$. 
The galaxies simply appear blue in the optical because the 
4000~\AA\ break has moved beyond the $I$ band. 
We illustrate this in Figure~\ref{fig10}b with a
$B-I$ versus $I-HK'$ plot. Here the solid (open) squares denote
the $I=20-23$ ($I=23-26$) sources from Figure~\ref{fig10}a.
Color-color tracks for an evolved
elliptical galaxy ({\it upper track}), Sb galaxy
({\it middle track}), and irregular galaxy ({\it bottom track})
from CWW are plotted as solid lines for $z<1$ and
as dashed lines for $z>1$. The figure shows that $z=1$
is a natural dividing line between the optically bright X-ray
sources, which are redder in $B-I$, and the optically faint X-ray
sources, which are bluer in $B-I$.

Finally, in Figure~\ref{fig10}c we plot $B-I$ versus redshift for
the sources with spectroscopic ({\it solid squares}) or
photometric ({\it open triangles}) redshifts to show that we 
have both photometric and some spectroscopic redshift confirmation 
for our interpretation. The sources with optically faint magnitudes 
$I=23-26$ are denoted by the second, larger symbols.

%
%
\begin{inlinefigure}
\psfig{figure=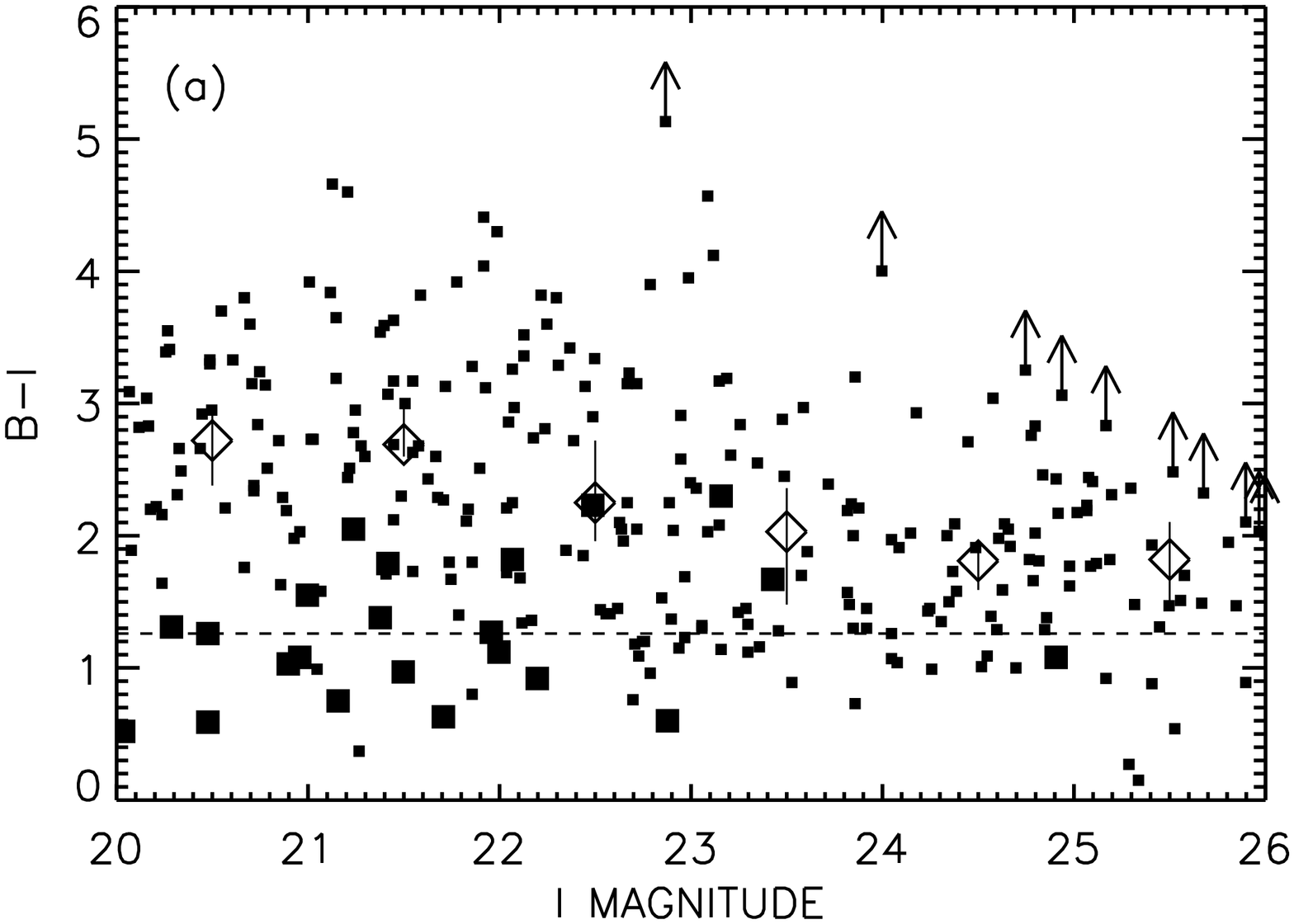,angle=90,width=3.3in}
\psfig{figure=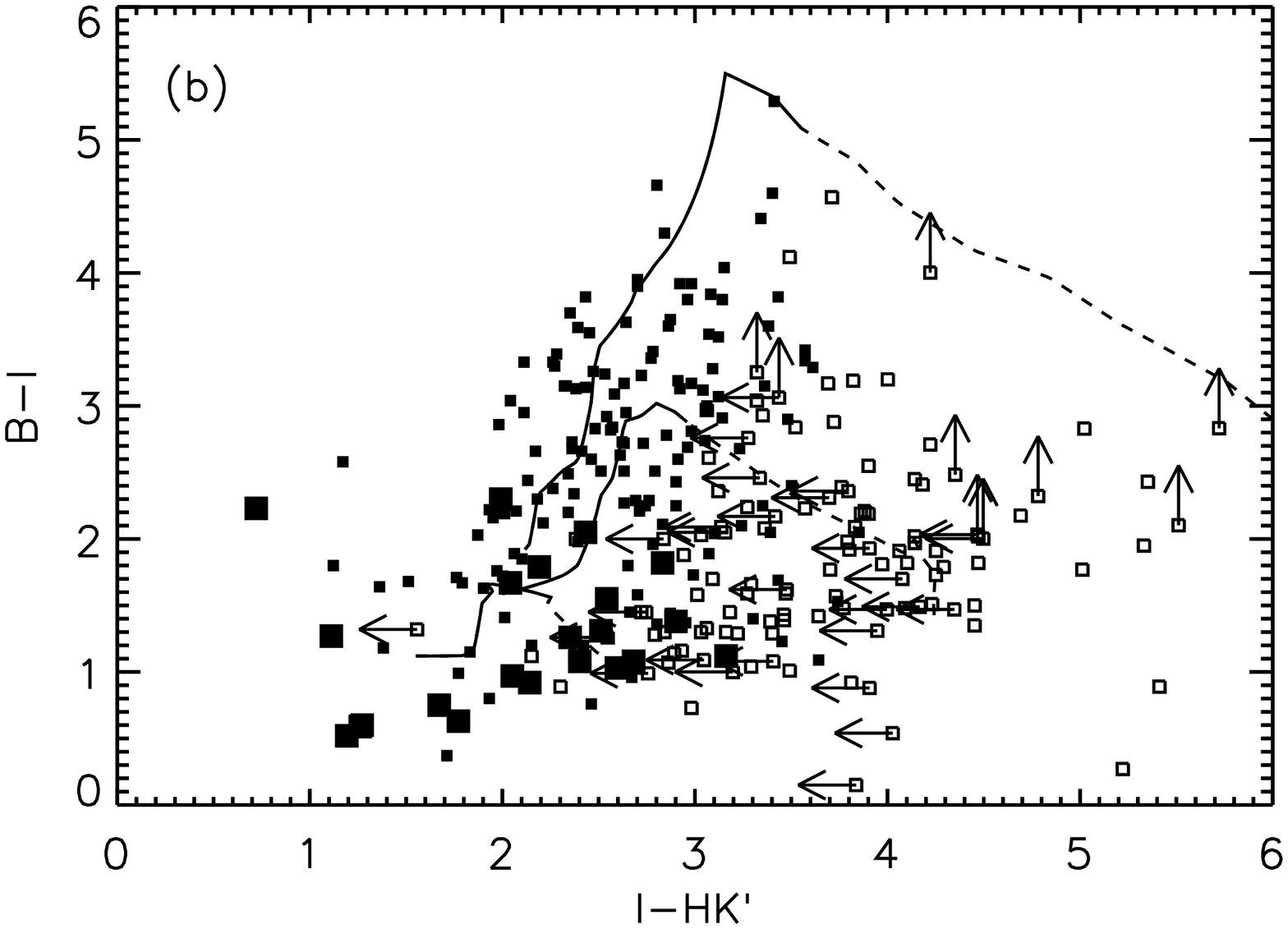,angle=90,width=3.3in}
\psfig{figure=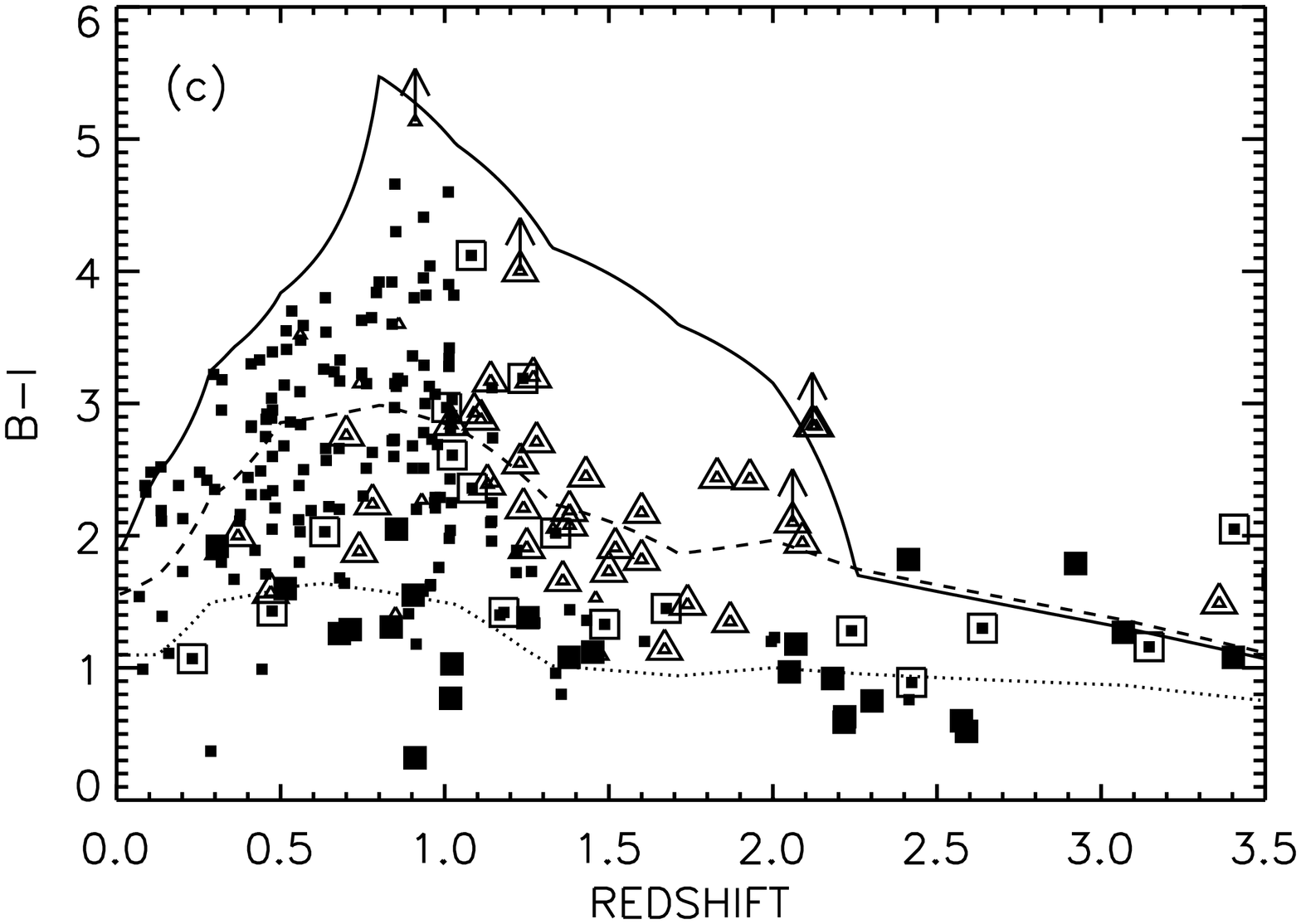,angle=90,width=3.3in}
\vspace{6pt}
\figurenum{10}
\caption{
(a) $B-I$ color vs. $I$ magnitude for the CDF-N X-ray sources with
X-ray positional uncertainties and optical counterpart separations
$<1''$ ({\it solid squares}). Sources undetected at the $2\sigma$
limit of $B=28.0$ are plotted at this magnitude with upward pointing
arrows. Large open diamonds with uncertainties show the median
$B-I$ colors of the non-broad-line sources in each magnitude bin. 
Dashed line shows the median color of all the broad-line AGNs.
(b) $B-I$ vs. $I-HK'$ color for the sources
from (a) with $I=20-23$ ({\it solid squares}) and
$I=23-26$ ({\it open squares}). Sources undetected at the $2\sigma$
limit of $B=28.0$ ($HK'=21.5$) are plotted at this magnitude with
upward (leftward) pointing arrows. Galaxy tracks for an evolved E
({\it upper track}), Sb ({\it middle track}), and
Irr ({\it bottom track}) from CWW are shown as solid
lines for $z<1$ and as dashed lines for $z>1$.
(c) $B-I$ vs. redshift for the sources from (b) with
spectroscopic ({\it solid squares}) or photometric 
({\it open triangles}) redshifts. Sources with $I=23-26$ are denoted
by a second, larger symbol. Sources undetected at the $2\sigma$
limit of $B=28.0$ are plotted at this magnitude with
upward pointing arrows. Evolved galaxy tracks are shown as solid 
(E), dashed (Sb), and dotted (Irr) lines. In all three panels 
broad-line sources are denoted by large solid squares.
\label{fig10}
}
\addtolength{\baselineskip}{10pt}
\end{inlinefigure}

%
%
\begin{inlinefigure}
\psfig{figure=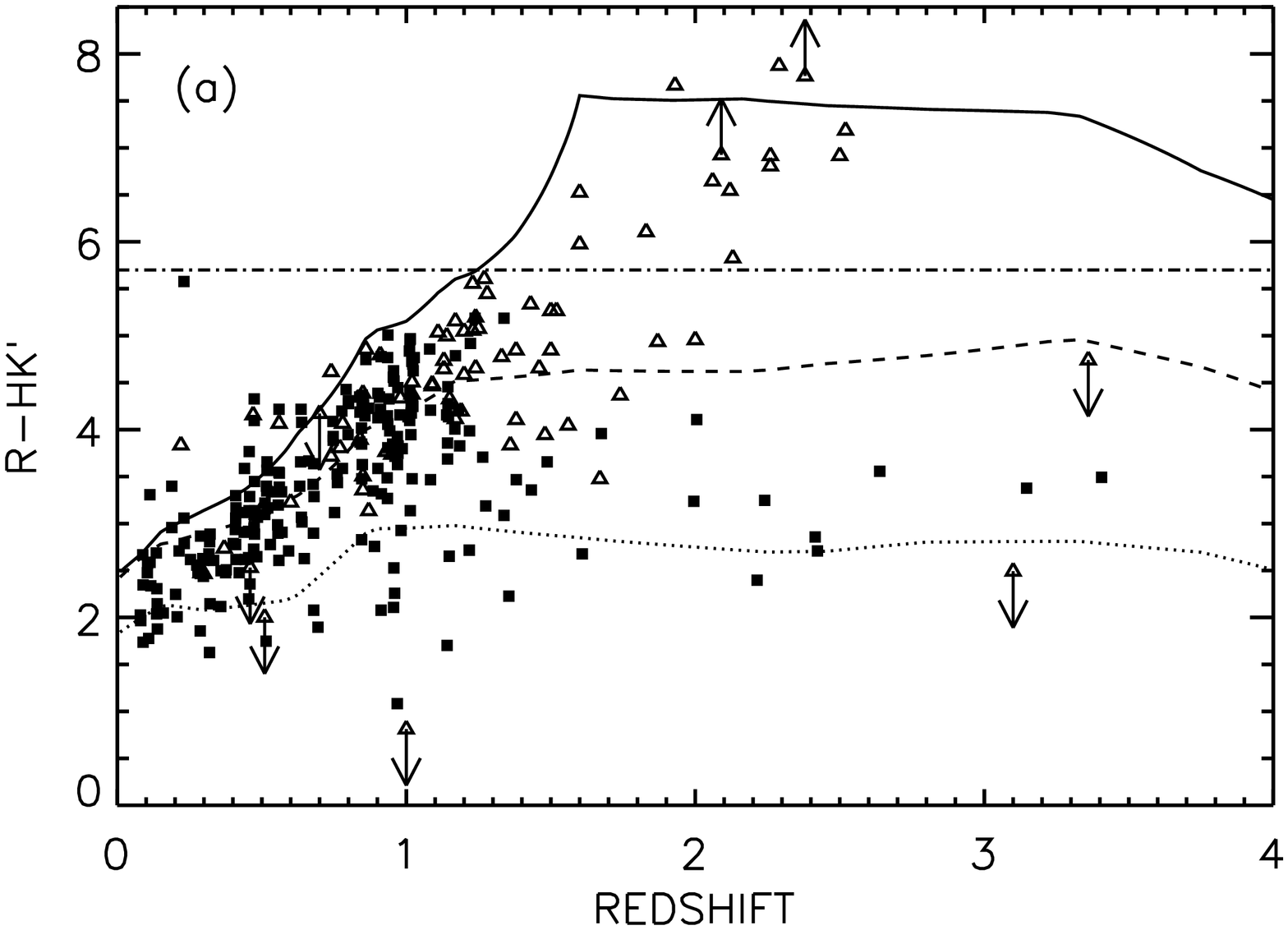,angle=90,width=3.5in}
\psfig{figure=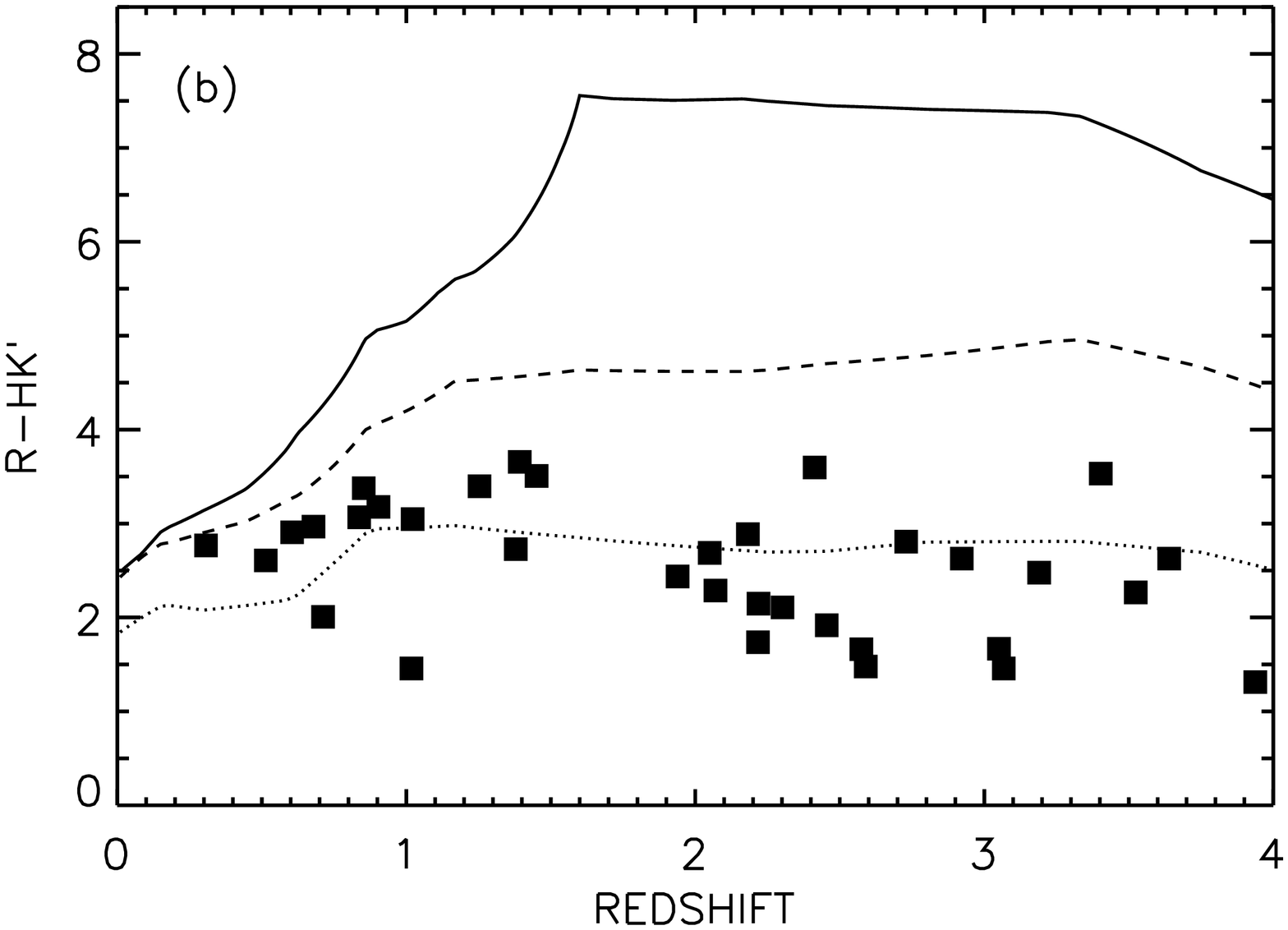,angle=90,width=3.5in}
\vspace{6pt}
\figurenum{11}
\caption{
$R-HK'$ color vs. redshift for the CDF-N X-ray sources with
(a) non-broad-line spectroscopic ({\it solid squares}) or
photometric ({\it open triangles}) redshifts and (b)
broad lines ({\it large solid squares}). Overlaid curves
show the galaxy redshift tracks for an evolved E ({\it solid}),
Sb ({\it dashed}), and Irr ({\it dotted}) with spectral shapes 
from CWW.
\label{fig11}
}
\addtolength{\baselineskip}{10pt}
\end{inlinefigure}

In fact, X-ray sources may be effective markers
of {\it red} galaxies at high redshifts, as suggested 
(and spectroscopically confirmed for one source at $z=1.467$
from near-infrared spectroscopy) by 
\markcite{cowie01}Cowie et al.\ (2001).
In Figure~\ref{fig11} we plot $R-HK'$ color versus redshift
for the CDF-N X-ray sources with (a) non-broad-line spectroscopic
({\it solid squares}) or photometric ({\it open triangles})
redshifts, and (b) broad-lines ({\it large solid squares}).
The overlays are CWW galaxy redshift
tracks for an evolved elliptical galaxy ({\it solid curve}),
Sb galaxy ({\it dashed curve}), and irregular galaxy
({\it dotted curve}). Sources with photometric redshifts
smoothly extend the spectroscopically identified source
population to higher redshifts and redder colors.
The upper envelope of the photometrically identified sources 
tracks the elliptical galaxy curve, and at the
higher redshifts, many of the galaxies fall into the Extremely
Red Object (ERO) color range (see also
\markcite{alex02b}Alexander et al.\ 2002b and
\markcite{mainieri02}Mainieri et al.\ 2002).
Here we conservatively define EROs
as having $R-HK'>5.7$, which is equivalent to $R-K>6$
($HK'-K\approx 0.3$; \markcite{barger99}Barger et al.\ 1999).
Just as for field EROs, these X-ray selected EROs
could be explained by either old stellar populations at $z>1$
or by dust-enshrouded galaxies. Most of the EROs in the CDF-N
sample ({\it circles} in Figure~\ref{fig2}) have
high X-ray-to-optical flux ratios because these sources
are faint in the optical, and perhaps this signature could be
used to pick out the higher redshift X-ray sources.
Although we do not have spectroscopic redshifts for any of
the sources redder than an Sb galaxy, and consequently our
photometric redshift estimates are unconfirmed, spectroscopic
identifications of high-redshift ($z>2$) red galaxies have recently 
been made, and the sources have been interpreted as the evolved 
descendants of galaxies that started forming stars at redshifts $z>4$
(\markcite{franx03}Franx et al.\ 2003;
\markcite{vd03}van Dokkum et al.\ 2003).

%
%
\begin{inlinefigure}
\psfig{figure=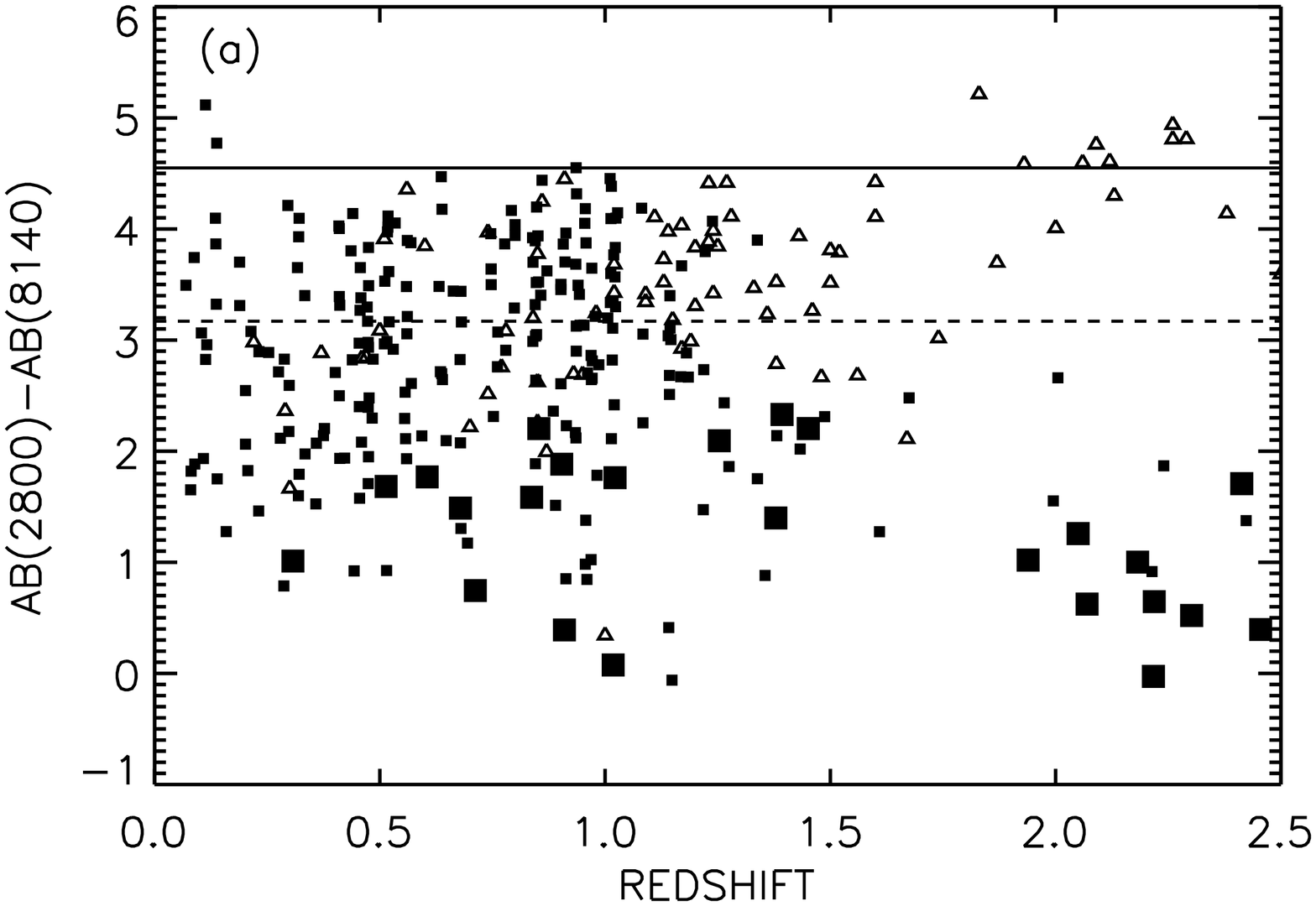,angle=90,width=3.5in}
\psfig{figure=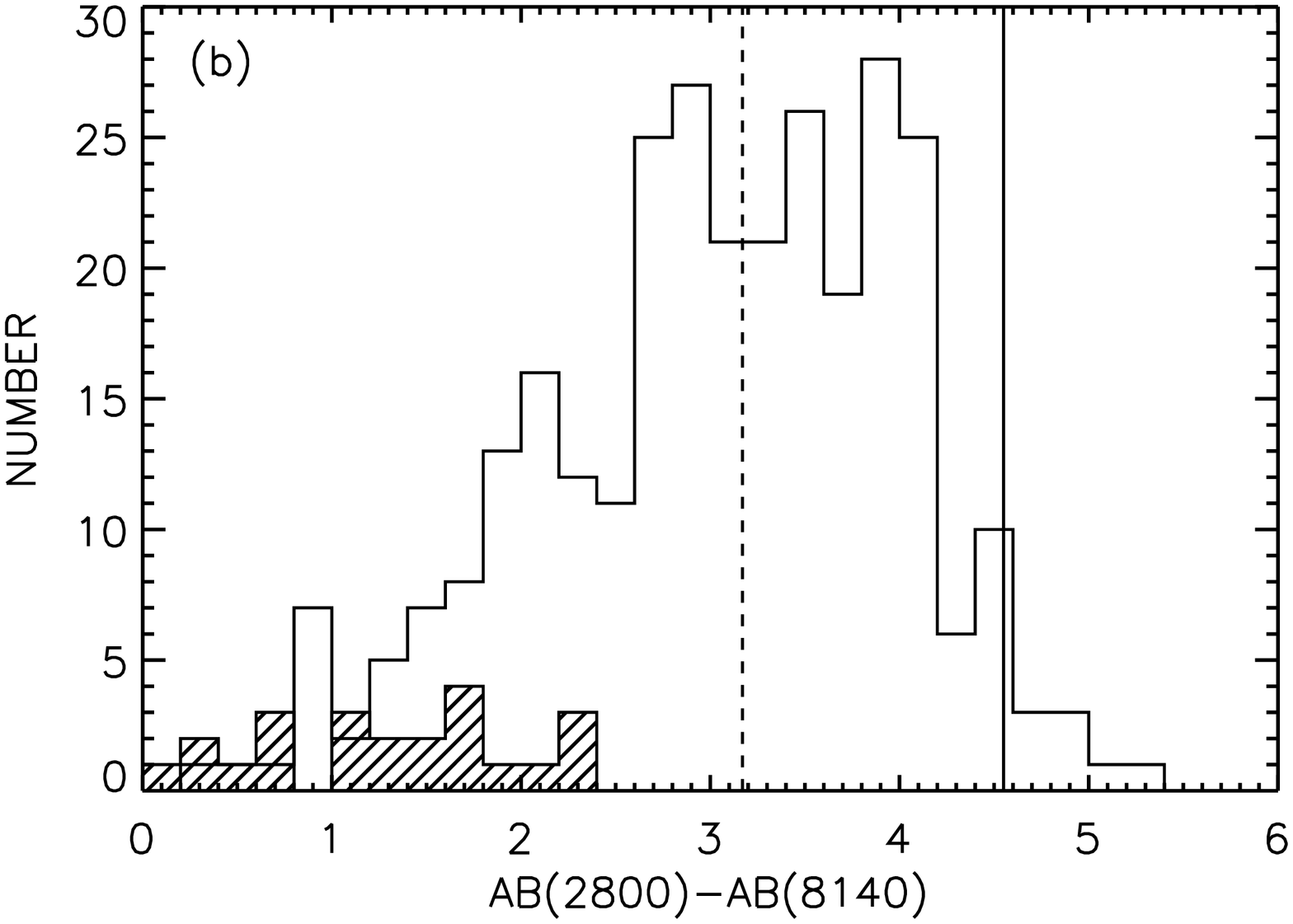,angle=90,width=3.5in}
\vspace{6pt}
\figurenum{12}
\caption{
(a) Rest-frame AB $2800-8140$~\AA\ color vs. redshift for
broad-line AGNs ({\it large solid squares}) and for the other
X-ray sources with spectroscopic ({\it solid squares}) or
photometric ({\it open triangles}) redshifts in the CDF-N.
Horizontal lines show the rest-frame colors of an E galaxy
({\it solid}) and an Sb galaxy ({\it dashed}).
Beyond $z\sim 1.5$ the  AB $8140$~\AA\ magnitudes are 
extrapolated and may be more uncertain.
(b) Rest-frame AB $2800-8140$~\AA\ color distributions for
$z<2.5$ broad-line AGNs ({\it hatched}) and other $z<2.5$ 
X-ray sources with spectroscopic or photometric redshifts 
({\it open}). Vertical lines show the rest-frame colors
of an E galaxy ({\it solid}) and an Sb galaxy ({\it dashed}).
\label{fig12}
}
\addtolength{\baselineskip}{10pt}
\end{inlinefigure}

It would be even more interesting to see how the {\it rest-frame} 
colors of the X-ray counterparts compare, and whether 
there are any intrinsic differences in the source properties 
between the brighter low-redshift sources and the fainter 
high-redshift ones. With our wide wavelength coverage, we are able 
to determine the rest-frame colors for the CDF-N X-ray sources with
redshift identifications.
In Figure~\ref{fig12}a we show rest-frame AB $2800-8140$~\AA\ color
versus redshift for the X-ray sources with either spectroscopic
or photometric redshifts. Beyond $z\sim 1.5$ the colors are
extrapolated and may be more uncertain.
The horizontal lines show the colors of an
elliptical galaxy ({\it solid}) and an Sb galaxy
({\it dashed}). The figure shows that many of the 
X-ray sources have the colors of evolved red galaxies and that 
there does not appear to be much evolution in the intrinsic colors 
of the galaxies with redshift. This is a rather surprising result, 
since passive evolution of the host galaxies would be expected 
to make the X-ray sources at high redshifts considerably bluer.

In Figure~\ref{fig12}b we show the AB $2800-8140$~\AA\ color 
distribution for the $z<2.5$ non-broad-line sources ({\it open}).
The vertical lines show the rest-frame
colors of an elliptical galaxy ({\it solid}) and an Sb
galaxy ({\it dashed}). Here we can more clearly see that a very 
large fraction of the X-ray population has rest-frame colors 
within a narrow red color range. In contrast, the $z<2.5$ 
broad-line sources ({\it hatched}) are very blue. This latter
result is expected, since the presence of a central AGN should 
make the rest-frame colors bluer than those of evolved galaxies.

The AGN contribution to the light is particularly strong in the 
ultraviolet and hence should increase the 2800~\AA\ flux relative 
to the 8140~\AA\ flux. However, this effect does not seem to be 
occurring in the non-broad-line sources. In fact, the number of 
sources with intermediate colors is rather small.
The situation therefore seems to be that either the AGNs
dominate the colors, as in the case of the broad-line sources,
or the AGNs do not have much effect on the colors, leaving
the sources to have the colors of evolved red galaxies.

%
%
\begin{inlinefigure}
\psfig{figure=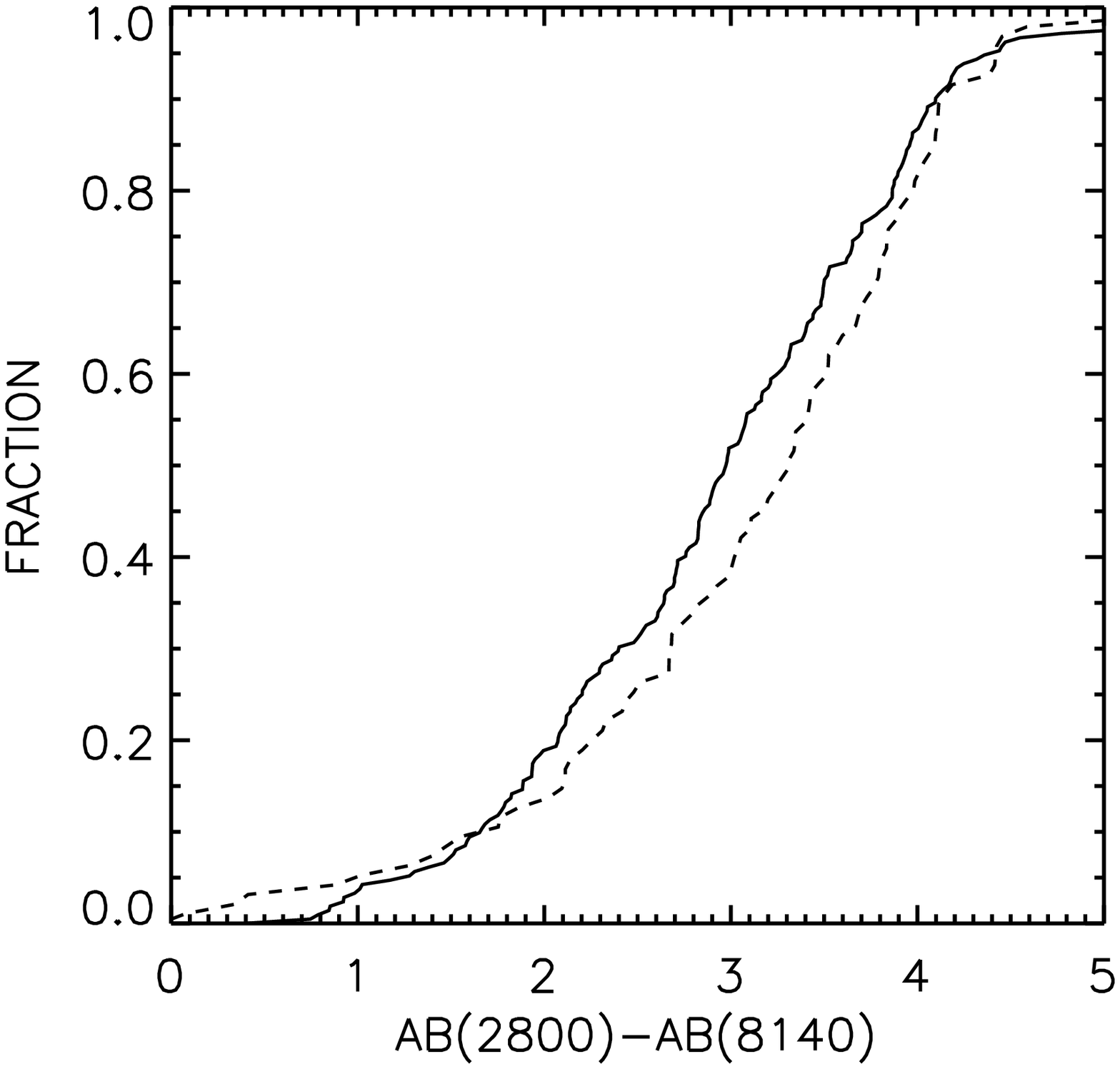,angle=90,width=3.5in}
\vspace{6pt}
\figurenum{13}
\caption{
Rest-frame AB $2800-8140$~\AA\ color distribution for the
CDF-N X-ray sources with either spectroscopic
or photometric redshifts in two redshift bins,
$z=0-1$ ({\it solid}) and $z=1-2$ ({\it dashed}).
\label{fig13}
}
\addtolength{\baselineskip}{10pt}
\end{inlinefigure}

We can examine the evolution of the rest-frame colors of the 
X-ray sources with either spectroscopic or photometric redshifts
more quantitatively by dividing the sample 
into two redshift bins, $z=0-1$ and $z=1-2$.
In Figure~\ref{fig13} we show the rest-frame AB $2800-8140$~\AA\ 
color distributions. There are 212 sources
in the $z=0-1$ ({\it solid}) interval and 95 in the $z=1-2$ 
({\it dashed}) interval. A Kolmogorov-Smirnov test 
shows that the maximum deviation of 0.14 is consistent with the 
two distributions being the same with a 95\% confidence criterion. 
The median color of the $z=0-1$ sources is 3.0,
and the median color of the $z=1-2$ sources is 3.3.

%
%
\begin{inlinefigure}
\psfig{figure=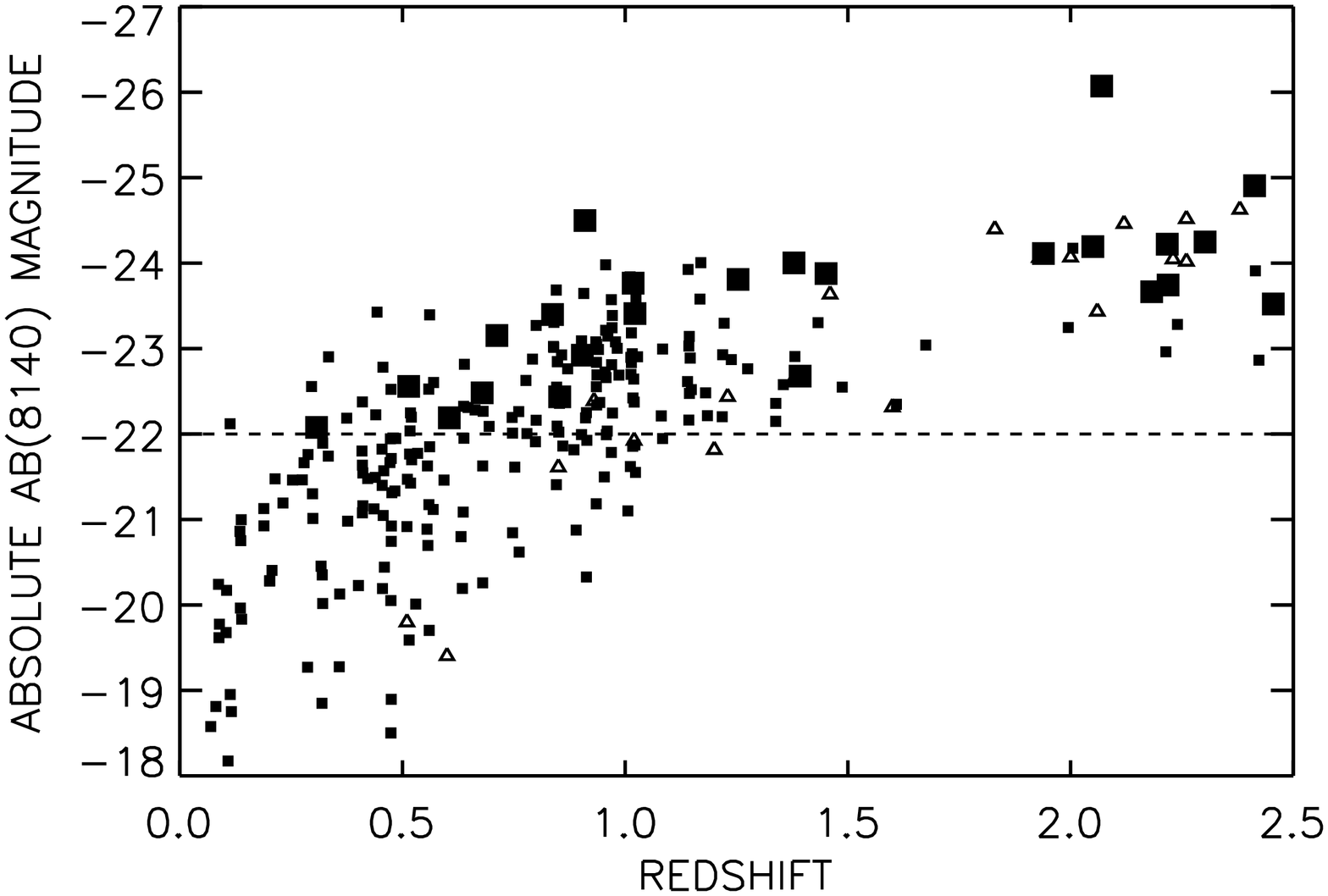,angle=90,width=3.5in}
\vspace{6pt}
\figurenum{14}
\caption{
Absolute AB 8140~\AA\ magnitude vs. redshift for the
spectroscopically identified non-broad-line sources
({\it solid squares}), photometrically identified
sources ({\it open triangles}), and broad-line AGNs
({\it large solid squares}) in the CDF-N. Dashed line
shows rough position of $M_I^\ast$.
\label{fig14}
}
\addtolength{\baselineskip}{10pt}
\end{inlinefigure}

One of the most striking features of Figure~\ref{fig10}b
is that there is a clear color-color separation
as a function of apparent magnitude. We might ask why apparent
magnitude seems to be a reasonably good redshift diagnostic for
the X-ray sources. The most likely answer
is that most of the host galaxies of the X-ray sources have
similar absolute magnitudes, and the X-ray sources lie
in a rather uniform host galaxy population. We examine this possibility
in Figure~\ref{fig14}, where we plot absolute AB 8140~\AA\
magnitude versus redshift for the spectroscopically identified
non-broad-line sources ({\it solid squares}),
photometrically identified sources ({\it open triangles}), and
broad-line AGNs ({\it large solid squares}). The bottom envelope
is just a selection effect against very faint optical magnitudes.
The dashed line shows the $M_B^\ast=-20.4$ galaxy from
\markcite{loveday92}Loveday et al.\ (1992), which roughly
corresponds to an Sb galaxy with $M_I^\ast=-22.0$.
From Figure~\ref{fig14} we see that many of the non-broad-line
sources are very luminous ($>M_I^\ast$), even at high redshifts. 
Thus, X-ray surveys may 
provide an effective way to find galaxies more luminous than $M^\ast$  
with colors similar to present-day early-type galaxies to high 
redshifts, though at present this interpretation is based on 
photometric redshift estimates that currently have relatively 
little confirmation at these redshifts.

\section{Summary}
\label{secsummary}

1.~We presented a catalog of optical and near-infrared magnitudes 
and optical spectroscopy for the X-ray sources identified in the 
$\approx 2$~Ms {\it Chandra} exposure of the Hubble Deep 
Field-North region. We now have redshifts for 284 of the 503 
X-ray point sources. The redshift identifications are very 
complete (87\%) for the $R\le 24$ sources. 

2.~We also presented photometric redshifts, obtained using seven
broadband colors and a Bayesian photometric redshift estimation
code, for 78 spectroscopically unidentified sources. 
We expect these photometric redshift estimates to be reliable
since we found that the photometric redshifts of all but 11 of 
the 172 spectroscopically identified non-broad-line sources with 
photometric redshift estimates are within 25\% of the spectroscopic 
redshifts. Since broad-line AGNs are straightforward to identify
spectroscopically, and we have now observed nearly all the X-ray 
sources, it is unlikely that broad-line AGNs are contaminating 
our photometric redshifts. 

3.~We showed that 
X-ray-to-optical flux ratios continue to be a good discriminator
between X-ray source classes down to very faint optical magnitudes 
and X-ray fluxes, with AGNs typically falling within the region
defined by the loci $\log(f_X/f_R)=\pm 1$ and quiescent galaxies
lying at $\log(f_X/f_R)\lesssim -2$.

4.~We constructed redshift slices of the hard X-ray sources 
versus hard X-ray flux and found 
that the spectroscopically identified sources already comprise 75\% 
of the measured light, of which 54\% arises from $z<1$ sources
and 68\% from $z<2$ sources. Our spectroscopic evidence for two 
broad redshift spikes in the X-ray source population discovered by 
\markcite{barger02}Barger et al.\ (2002) has continued to grow,
illustrating that the CDF-N is dominated by a rather small number 
of independent velocity elements.
Because of the apparent velocity sheets and complex structures 
in the redshift distributions of the X-ray sources in individual
{\it Chandra} fields, we will need to observe a large number of fields
in order to get the average true redshift distribution.

5.~We found that X-ray sources may be effective markers of
red galaxies at high redshifts. In an $R-HK'$ versus 
redshift plot, the sources with photometric redshifts smoothly 
extend the spectroscopically identified source population to 
higher redshifts and redder colors. The upper envelope of the 
photometrically identified sources follows that of an elliptical 
galaxy track, and fourteen of these sources have $R-HK'$ colors 
that would classify them as EROs with redshifts between 
$z\sim 1.5$ and $z\sim 2.5$.

6.~With our wide wavelength coverage, we were also able to 
determine rest-frame colors for the X-ray sources with redshift
identifications. We found that many of the non-broad-line X-ray 
sources have intrinsic colors of evolved bulge-dominated galaxies 
and that there is very little evolution in these colors with redshift, 
as confirmed with a Kolmogorov-Smirnov test between two redshift
bins, $z=0-1$ and $z=1-2$. If anything, the galaxies in the
higher redshift interval are redder than those in the lower 
redshift interval.
This is surprising both because passive evolution of the host
galaxies would be expected to make the X-ray sources at higher
redshifts bluer and because any AGN contributions to the light
would make the sources bluer. Thus, it seems
that either the AGN light dominates the colors, as is the case 
for the broad-line sources, or the AGN presence does
not have much effect on the colors.

7.~We showed that apparent magnitude seems to be a reasonably good 
redshift diagnostic for the non-broad-line X-ray sources, with 
there being a clear separation at a redshift of $z\sim 1$ between 
optically bright and optically faint sources in a $B-I$ versus 
$I-HK'$ color-color plot. 
We suggest that this results from the 
fact that the X-ray sources lie in a rather uniform host galaxy
population with similar absolute magnitudes. We found that many 
of the non-broad-line sources are very luminous ($>M_I^\ast$),
even at high redshifts. We therefore infer that X-ray surveys may 
provide an effective way to find $M^\ast$ galaxies with colors 
similar to present-day early-type galaxies to high redshifts,
though at present this interpretation is based on photometric 
redshift estimates that currently have relatively little 
confirmation at these redshift values.

\acknowledgements
We thank the referee, David Helfand, for helpful comments
that improved the manuscript.
We are grateful to Herv{\'e} Aussel for providing 
supplemental optical spectra.
We gratefully acknowledge support from the University of Wisconsin 
Research Committee with funds granted by the Wisconsin Alumni 
Research Foundation (A.~J.~B.), the Alfred P. Sloan Foundation
(A.~J.~B.), NSF grants AST-0084847 (A.~J.~B), 
AST-0084816 (L.~L.~C.), and AST 99-83783 (D.~M.~A., W.~N.~B.),
and NASA grants DF1-2001X (L.~L.~C.),
GO2-3187B (L.~L.~C.), NAS 8-01128 (G.~P.~G.), and
G02-3187A (D.~M.~A., F.~E.~B., W.~N.~B.).

\appendix
In Table~A1 we present optical magnitudes and spectroscopic 
measurements, where available, for the full X-ray point source 
catalog of A03. The 503 X-ray sources are identified by A03 
number (col.~[1]) and by right ascension and declination
coordinates in decimal degrees (cols.~[2] and [3]).
We also give right ascension and declination coordinates where
the aperture magnitudes were measured (cols.~[4] and [5]).
In columns~$(6)-(7)$ we give the soft and hard X-ray fluxes
from A03 in units of $10^{-15}$~ergs~cm$^{-2}$~s$^{-1}$. 
In columns~$(8)-(17)$ we give
isophotal $R$ magnitudes, $HK'$, $z'$, $I$, $R$, $V$,
$B$, and $U$ corrected-aperture magnitudes, spectroscopic 
redshifts, and photometric redshifts. 
Objects where a spectrum was obtained but no identification
could be made are marked ``Obs'' for observed in the spectroscopic
redshift column (col.~[16]). The final column contains a ``B''
if the spectrum has broad emission lines, an ``s'' if the spectroscopic
identification is at all suspect, and an ``m'' if the source is complex
or if a bright nearby neighbor may be contaminating the photometry.

Source 267 is more likely associated with an extremely
red object (\markcite{dickinson00}Dickinson et al.\ 2000)
to the northeast of the $z=0.401$ galaxy identified here,
but the adopted automatic procedure gives the $z=0.401$
galaxy as the counterpart. We note that 
V. L. Sarajedini (priv. comm.) has identified the $z=0.401$ galaxy
as variable, which may suggest that some part of the X-ray 
emission may also be coming from this object.

Counterparts or redshifts for five new sources differ from 
the values given in the \markcite{barger02}Barger et al.\ (2002) 
table:

$\bullet$~Source 91: The new X-ray position has moved the source 
south of the southern component of this merging pair, whose 
$z=0.294$ redshift was given as the redshift for
source 71 in Barger et al. The X-ray source now lies 
on the tidal tail. It appears likely that the $z=0.294$ redshift 
is still correct, but the positional offset means that it no longer 
meets the formal criteria for inclusion in the table.

$\bullet$~Sources 119 and 120: Source 92 of Barger et al.\ has 
now split into these two sources. Source 119 has the redshift
of the Barger et al.\ source 92 ($z=0.473$), while source 120 
has a redshift of $z=0.695$.

$\bullet$~Source 259: In Barger et al.\ the redshift for this 
source (their source 184) was given as $z=2.315$. Newer spectra
show this redshift was based on an incorrect identification 
of an emission line as Ly$\alpha$, when it is really CIV.
The present identification is based on [OII]~3727~\AA,
MgII~2800~\AA, CIII]~1909~\AA, and CIV~1550~\AA\ and appears 
unambiguous.

$\bullet$~Source 443: This source has two possible counterparts, 
a brighter one at $z=0.033$ and a fainter one at $z=0.231$.
We have adopted the $z=0.231$ redshift of the closer source
instead of the $z=0.033$ redshift given in Barger et al.\ 
(their source 331).

The higher resolution images were smoothed to match the seeing in
the lower resolution images in order to give the most precise
color measurements. The fluxes for each source were measured
using $3''$ diameter apertures on the smoothed images.
The Suprime-Cam photometry was scaled to match the
HDF-N proper photometry of \markcite{fernandez99}Fern{\'a}ndez-Soto,
Lanzetta, \& Yahil (1999), both for absolute calibration and to
approximate total magnitudes (the Fern{\'a}ndez-Soto et al.
measurements are isophotal). The $HK'$ magnitudes were corrected
to approximate total magnitudes using an offset calculated
as the median difference between $3''$ and $6''$ aperture
magnitudes measured for a complete $HK'<20.5$ sample in the field.

We give the $B$ and $V$ magnitudes in the Johnson system,
the $R$ and $I$ magnitudes in the Kron-Cousins system,
the $HK'$ magnitudes in the Wainscoat-Cowie system
(\markcite{wains92}Wainscoat \& Cowie 1992), and
the $U$ and $z'$ magnitudes in the AB magnitude system.
Offsets of $-0.077$, 0.023, 0.228, 0.453, and 1.595
for the $B$, $V$, $R$, $I$, and $HK'$ bands, respectively,
can be added to the magnitudes in Table~A1 to get AB magnitudes.
Saturation occurs in the images at magnitudes of about 16 ($U$),
20.6 ($B$), 19.8 ($V$), 20 ($R$), 19.7 ($I$), 19.4 ($z'$),
and 14.4 ($HK'$). Magnitudes brighter than these
values in Table~A1 are likely to be underestimated.
Sources that had a negative flux in the aperture are listed 
with a negative magnitude; in these cases, the absolute value 
of the magnitude was computed from the absolute value of the flux.
If no magnitude is given, then this indicates that the magnitude
was severely contaminated, usually by a nearby bright star.

We show $z'$ band thumbnail images of the CDF-N sources in
Figure~A1, ordered by A03 catalog number. We use the 
$z'$ band because many of the sources are red and hence
more easily seen in this band. Each thumbnail is $18''$ on
a side. The expected X-ray position of each source is shown
with a $4''$ diameter circle (representing the $2''$ search
radius), and the position where the $3''$ diameter aperture
magnitude was measured is shown with a $3''$ diameter circle.
The number of the source from Table~A1 is printed in
the top left corner. If there is a spectroscopic identification,
then the redshift (or the word ``STAR'' for a stellar spectrum)
is printed in the bottom left corner. If a nearby faint source 
and a more distant bright source with a spectroscopic
redshift are both found to lie within the $2''$ search radius,
then we have identified the X-ray source with the faint
source and zeroed out the spectroscopic redshift.
If there is a spectroscopic identification for a source that
lies outside our chosen cross-identification radius of $2''$
but still within a $5''$ radius about the X-ray position, then
the redshift of that source is printed in fainter text to indicate
that the redshift has been given for interest but has not been
used. The only exceptions are the seven off-axis X-ray sources
discussed in \S~\ref{secimaging}. If there is no spectroscopic
redshift for a source, but there is a photometric redshift 
estimate, then the photometric redshift estimate is printed in 
parentheses.


\newpage

\begin{references}

\reference{akiyama00}
Akiyama, M., et al.\ 2000, \apj, 532, 700

\reference{alex02a}
Alexander, D. M., Aussel, H., Bauer, F. E., Brandt, W. N.,
Hornschemeier, A. E., Vignali, C., Garmire, G. P., \&
Schneider, D. P.\ 2002a, \apj, 568, L85

\reference{alex02b}
Alexander, D. M., Vignali, C., Bauer, F. E., Brandt, W. N., 
Hornschemeier, A. E., Garmire, G. P., \& Schneider, D. P.\ 2002b,
\aj, 123, 1149

\reference{alex03}
Alexander, D. M., et al.\ 2003, \aj, in press (astro-ph/0304392) (A03)

\reference{barden94}
Barden, S. C., Armandroff, T., Muller, G., Rudeen, A. C., Lewis, J.,
\& Groves, L.\ 1994, Proc. SPIE, 2198, 87

\reference{barger99}
Barger, A. J., Cowie, L. L., Trentham, N., Fulton, E., Hu, E. M.,
Songaila, A., \& Hall, D.\ 1999, \aj, 117, 102

\reference{barger01a}
Barger, A. J., Cowie, L. L., Bautz, M. W., Brandt, W. N.,
Garmire, G. P., Hornschemeier, A. E., Ivison, R. J., \&
Owen, F. N.\ 2001a, \aj, 122, 2177

\reference{barger01b}
Barger, A. J., Cowie, L. L., Mushotzky, R. F., \&
Richards, E. A.\ 2001b, \aj, 121, 662

\reference{barger02}
Barger, A. J., Cowie, L. L., Brandt, W. N., Capak, P.,
Garmire, G. P., Hornschemeier, A. E., Steffen, A. T., \&
Wehner, E. H.\ 2002, \aj, 125, 1839

\reference{bauer02}
Bauer, F. E., Alexander, D. M., Brandt, W. N., Hornschemeier, A. E.,
Vignali, C., Garmire, G. P., \& Schneider, D. P.\ 2002, \aj, 124, 2351

\reference{benitez}
Ben{\'i}tez, N.\ 2000, \apj, 536, 571

\reference{bol00}
Bolzonella, M., Miralles, J.-M., \& Pell{\'o}, R.\ 2000, A\&A, 363, 467

\reference{brandt01}
Brandt, W. N., et al.\ 2001, \aj, 122, 2810

\reference{bc93}
Bruzual, A. G. \& Charlot, S.\ 1993, \apj, 405, 538

\reference{cabanac}
Cabanac, R. A., de Lapparent, V., \& Hickson, P.\ 2002,
A\&A, 389, 1090

\reference{capak03a}
Capak, P., et al.\ 2003a, \aj, submitted

\reference{capak03b}
Capak, P., et al.\ 2003b, in preparation

\reference{cohen00}
Cohen, J. G., Hogg, D. W., Blandford, R., Cowie, L. L., Hu, E.,
Songaila, A., Shopbell, P., \& Richberg, K.\ 2000, \apj, 538, 29

\reference{cww}
Coleman, G. D., Wu, C-C., \& Weedman, D. W.\ 1980, \apjs, 43, 393 (CWW)

\reference{comastri02}
Comastri, A., et al.\ 2002, \apj, 571, 771

\reference{connolly}
Connolly, A. J., Csabai, I., Szalay, A. S., Koo, D. C.,
Kron, R. G., \& Munn, J. A.\ 1995, \aj, 110, 2655

\reference{cowie96}
Cowie, L. L., Songaila, A., Hu, E. M., \& Cohen, J. G.\ 1996,
\aj, 112, 839

\reference{cowie01}
Cowie, L. L., et al.\ 2001, \apj, 551, L9

\reference{cowie02}
Cowie, L. L., Garmire, G. P., Bautz, M. W., Barger, A. J.,
Brandt, W. N., \& Hornschemeier, A. E.\ 2002, \apj, 566, L5

\reference{cowie03}
Cowie, L. L., Barger, A. J., Bautz, M. W., Brandt, W. N., \&
Garmire, G. P.\ 2003, \apj, 584, L57

\reference{dickinson00}
Dickinson, M., et al.\ 2000, \apj, 531, 624

\reference{faber02}
Faber, S. M., et al.\ 2002, Proc. SPIE, 4841, 186

\reference{fernandez99}
Fern{\'a}ndez-Soto, A., Lanzetta, K. M., \& Yahil, A.\ 1999,
\apj, 523, 72

\reference{franx03}
Franx, M. et al.\ 2003, \apj, 587, L79

\reference{gehrels86}
Gehrels, N.\ 1986, ApJ, 303, 336

\reference{giacconi02}
Giacconi, R., et al.\ 2002, \apjs, 139, 369

\reference{gilli03}
Gilli, R., et al.\ 2003, \apj, in press (astro-ph/0304177)

\reference{hasinger02}
Hasinger, G.\ 2002, in New Visions of the X-ray Universe in the
{\it XMM-Newton} and {\it Chandra} Era, ed. F. Jansen,
(ESA SP-488; Noordwijk: ESA/ESTEC), in press (astro-ph/0202430)

\reference{hodapp96}
Hodapp, K.-W., et al.\ 1996, NewA, 1, 177

\reference{horn01}
Hornschemeier, A. E., et al.\ 2001, \apj, 554, 742

\reference{horn03}
Hornschemeier, A. E., et al.\ 2003, \aj, in press 
(astro-ph/0305086)

\reference{jacoby98}
Jacoby, G. H., Liang, M., Vaughnn, D., Reed, R., \&
Armandroff, T.\ 1998, Proc. SPIE, 3355, 721

\reference{kinney}
Kinney, A. L., Calzetti, D., Bohlin, R. C., McQuade, K.,
Storchi-Bergmann, T., \& Schmitt, H. R.\ 1996, \apj, 467, 38

\reference{koekemoer02}
Koekemoer, A. M., et al.\ 2002, \apj, 567, 657

\reference{lehmann}
Lehmann, I., et al.\ 2001, A\&A, 371, 833

\reference{loveday92}
Loveday, J., Peterson, B. A., Efstathiou, G., \& Maddox, S. J.\ 1992,
\apj, 390, 338

\reference{maccacaro88}
Maccacaro, T., Gioia, I. M., Wolter, A., Zamorani, G., \&
Stocke, J. T.\ 1988, \apj, 326, 680

\reference{mainieri02}
Mainieri, V., Bergeron, J., Hasinger, G., Lehmann, I., Rosati, P.,
Schmidt, M., Szokoly, G., \& Della Ceca, R.\ 2002, A\&A, 393, 425

\reference{mcmahon02}
McMahon, R. G., White, R. L., Helfand, D. J., \& Becker, R. H.\ 2002,
\apjs, 143, 1

\reference{miyazaki02}
Miyazaki, S., et al.\ 2002, \pasj, 54, 833

\reference{moran02}
Moran, E. C., Filippenko, A. V., \& Chornock, R.\ 2002, \apj, 579, L71

\reference{muller98}
Muller, G. P., Reed, R., Armandroff, T., Boroson, T. A.,
\& Jacoby, G. H.\ 1998, Proc. SPIE, 3355, 577

\reference{oke95}
Oke, J.B., et al.\ 1995, \pasp, 107, 375

\reference{richards00}
Richards, E. A.\ 2000, \apj, 533, 611

\reference{rosati02}
Rosati, P., et al.\ 2002, \apj, 566, 667

\reference{schmidt98}
Schmidt, M., et al.\ 1998, A\&A, 329, 495

\reference{schreier01}
Schreier, E. J., et al.\ 2001, \apj, 560, 127

\reference{vd03}
van Dokkum, P. G., et al.\ 2003, \apj, 587, L83

\reference{wains92}
Wainscoat, R. J., \& Cowie, L. L.\ 1992, \aj, 103, 332

\reference{wolfe98}
Wolfe, T., Reed, R., Blouke, M. M., Boroson, T. A.,
Armandroff, T., \& Jacoby, G. H.\ 1998, Proc. SPIE, 3355, 487

\reference{yang03}
Yang, Y., Mushotzky, R. F., Barger, A. J., Cowie, L. L.,
Sanders, D. B., \& Steffen, A. T.\ 2003, \apj, 585, L85

\end{references}
\end{document}